

\newcount\instnumber  \instnumber=0
\def\instfoot#1{\global\advance\instnumber by1
  \xdef#1{$^{(\the\instnumber)}$}}
%
%
\catcode`@=11 
%
\def\space@ver#1{\let\@sf=\empty \ifmmode #1\else \ifhmode
   \edef\@sf{\spacefactor=\the\spacefactor}\unskip${}#1$\relax\fi\fi}
\newdimen\authrefminspace  \authrefminspace=2pc
\newcount\authrefcount     \authrefcount=96
\newif\ifauthreferenceopen       \newwrite\authreferencewrite
\newtoks\authrw@toks
\let\PRauthrefmark=\attach
\def\authrefmark#1{\relax\PRauthrefmark{#1}}
\newcount\lastauthrefsbegincount \lastauthrefsbegincount=0
\def\authrefsend{\authrefmark{\count255=\authrefcount
   \advance\count255 by-\lastauthrefsbegincount
   \ifcase\count255 \number\authrefcount
   \or \number\lastauthrefsbegincount,\number\authrefcount
   \else \number\lastauthrefsbegincount-\number\authrefcount \fi}}
\def\authrefch@ck{\chardef\authrw@write=\authreferencewrite
   \ifauthreferenceopen \else \authreferenceopentrue
   \immediate\openout\authreferencewrite=authrefe.texauxil \fi}
{\catcode`\^^M=\active 
  \gdef\obeyendofline{\catcode`\^^M\active \let^^M\ }}%
{\catcode`\^^M=\active 
  \gdef\ignoreendofline{\catcode`\^^M=5}}
{\obeyendofline\gdef\authrw@start#1{\def\t@st{#1} \ifx\t@st\blankend%
\endgroup \@sf \relax \else \ifx\t@st\bl@nkend \endgroup \@sf \relax%
\else \authrw@begin#1
\backtotext
\fi \fi } }
{\obeyendofline\gdef\authrw@begin#1
{\def\n@xt{#1}\authrw@toks={#1}\relax%
\authrw@next}}
\newif\iffirstauthrefline  \firstauthreflinetrue
\def\rwr@teswitch{\ifx\n@xt\blankend \let\n@xt=\authrw@begin %
 \else\iffirstauthrefline \global\firstauthreflinefalse%
\immediate\write\authrw@write{\noexpand\obeyendofline \the\authrw@toks}%
\let\n@xt=\authrw@begin%
      \else\ifx\n@xt\authrw@@d \def\n@xt{\immediate\write\authrw@write{%
        \noexpand\ignoreendofline}\endgroup \@sf}%
             \else \immediate\write\authrw@write{\the\authrw@toks}%
             \let\n@xt=\authrw@begin\fi\fi \fi}
\def\authrw@next{\rwr@teswitch\n@xt}
\def\authrw@@d{\backtotext} \let\authrw@end=\relax
\let\backtotext=\relax
\newdimen\authrefindent     \authrefindent=17pt
\def\authrefitem#1{\par \parskip=0pt
\hangafter=0 \hangindent=\authrefindent \Textindent{#1}}
\def\authrefnum#1{\space@ver{}\authrefch@ck \firstauthreflinetrue%
 \global\advance\authrefcount by 1 \xdef#1{\the\authrefcount}}

\def\authref#1{\authrefnum#1%
 \immediate\write\authreferencewrite{\noexpand\authrefitem{(\char#1)}}%
\begingroup\obeyendofline\authrw@start}
%

%
%
\catcode`@=12 
%
91  rep
 
\overfullrule=0pt
 \linepenalty=100
\uchyph=200
\brokenpenalty=200
\hbadness=10000
\clubpenalty=10000
\widowpenalty=10000
\displaywidowpenalty=10000
p
\pretolerance=10000
\tolerance=2000
\nobreak
\penalty 5000
\hyphenpenalty=5000
\exhyphenpenalty=5000

\font\eightrm=cmr8 scaled\magstep1
 
  \def\str{\penalty-10000\hfilneg\ } 
 
 
\Ref\QCD{H.~Fritzsch, M.~Gell-Mann and H.~Leutwyler, Phys.~Lett.
                    {\bf B47} (1973) 365,\hfil\break
         D.J.~Gross and F.~Wilczek, Phys.~Rev.~Lett. {\bf 30} (1973) 1343,
          \hfil\break
         H.D.~Politzer, Phys.~Rev.~Lett. {\bf 30} (1973) 1346.}
\Ref\JETR{SLD Collab., K.~Abe {\it et al.}, Phys.~Rev.~Lett. {\bf 71}
          (1993) 2528.}
\Ref\EEC{C.L.~Basham {\it et al.},
                      Phys.~Rev.~Lett. {\bf 41} (1978) 1585,
                      Phys.~Rev. {\bf D17} (1978) 2298,
                      Phys.~Rev. {\bf D19} (1979) 2018. }
\Ref\RSE{D.G.~Richards, W.J.~Stirling and S.D.~Ellis,
          Nucl.~Phys. {\bf B229} (1983) 317. }
\Ref\AB{A.~Ali and F.~Barreiro, Nucl.~Phys. {\bf B236} (1984) 269. }
\Ref\FK{N.K.~Falck and G.~Kramer, Z.~Phys. {\bf C42} (1989) 459.}
\Ref\KN{Z.~Kunszt and P.~Nason, Z Physics at LEP 1,
          CERN Report CERN 89--08, Vol.~1
          eds. G.~Altarelli, R. Kleiss and C.~Verzegnassi, p.~373. }
\Ref\PDB{Review of Particle Properties, K. Hikasa {\it et al.},
         Phys.~Rev. {\bf D45} (1992)
         III.54.}
\Ref\LTHREEa{L3 Collab., B.~Adeva {\it et al.},
                       Phys.~Lett. {\bf B257} (1991) 469. }
\Ref\ALEPHa{ALEPH Collab., D.~Decamp {\it et al.},
                       Phys.~Lett. {\bf B257} (1991) 479. }
\Ref\OPALa{OPAL Collab., P.D.~Acton {\it et al.},
                    Phys.~Lett. {\bf B276} (1992) 547. }
\Ref\NLL{S.~Catani {\it et al.}, Nucl.~Phys. {\bf B407} (1993) 3.}
\Ref\TURNOCK{G.~Turnock, Cambridge Preprint Cavendish--HEP--92/3 (1992).}
\Ref\FIORE{R.~Fiore, {\it et al.}, Phys.~Lett. {\bf B294} (1992) 431.}
\Ref\COLLINS{J.C.~Collins and D.E.~Soper, Nucl.~Phys. {\bf B193} (1981) 381.}
\Ref\OPAL{OPAL Collab., P.D.~Acton {\it et al.},
                       Z.~Phys. {\bf C59} (1993) 1. }
\Ref\WEBBER{B.R.~Webber, CERN--TH--6706--92 (1992).}
\Ref\DELPHIa{DELPHI Collab., P.~Abreu {\it et al.},
                       Phys.~Lett. {\bf B252} (1990) 149. }
\Ref\DELPHIb{DELPHI Collab., P.~Abreu {\it et al.},
                       Z.~Phys. {\bf C59} (1993) 21. }
\Ref\OPALb{OPAL Collab., M.Z.~Akrawy {\it et al.},
                    Phys.~Lett. {\bf B252} (1990) 159. }
\Ref\ALEPH{ALEPH Collab., D.~Decamp {\it et al.},
                       Phys.~Lett. {\bf B284} (1992) 163. }
\Ref\LTHREE{L3 Collab., O.~Adriani {\it et al.},
                       Phys.~Lett. {\bf B284} (1992) 471. }
\Ref\SLD{ SLD Design Report, SLAC Report 273 (1984). }
\Ref\VXD{ C.J.S.~Damerell {\it et al.}, Nucl.~Inst.~Meth.
          {\bf A288} (1990) 288.}
\Ref\LAC{ D.~Axen {\it et al.}, Nucl.~Inst.~Meth.
          {\bf A328} (1993) 472.}
\Ref\WIC{ A.C. Benvenuti {\it et al.},
          Nucl.\ Instr.\ Meth.\ {\bf A290} (1990) 353. }
\Ref\THRUST{E.~Farhi, Phys.~Rev.~Lett. {\bf 39} (1977) 1587.}
\Ref\JETSET{ T.~Sj\"{o}strand and M.~Bengtsson,
             Comp.~Phys.~Comm. {\bf 43} (1987) 367. }
\Ref\HERWIG{ G.~Marchesini {\it et al.},
             Comp.~Phys.~Commun. {\bf 67} (1992) 465. }
\Ref\GEANT{S.L.~Linn {\it et al.}, CERN--DD/78/12 (1978).}
\Ref\TASSO{ TASSO Collab, W.~Braunschweig {\it et al.},
            Z.~Phys., {\bf C41} (1988) 359;  \nextline
            P.~N.~Burrows, Z.~Phys. {\bf C41} (1988) 375. }
\Ref\OPALc{OPAL Collab., M.Z.~Akrawy {\it et al.},
                       Z.~Phys. {\bf C47} (1990) 505. }
\Ref\WEBBERb{ B.R.~Webber, private communications.}
\Ref\PNBHM{The criteria for convergence are discussed in:
P.N. Burrows and H. Masuda, SLAC--PUB--6394 (1993); to appear in Z. Phys. C.}
\def\lam{\Lambda_{\overline{MS}}}
\def\as{\alpha_s}
\def\asz{\alpha_s(M_Z^2)}
\def\IR{initial state photon radiation}
\def\MC{Monte Carlo}
\def\Oa2{{\cal O}(\alpha_s^2)}

\def\Rmatch{{\it R-matching}}
\def\mRmatch{{\it modified R-matching}}

\def\doeack{\footnote {\star}
{\eightrm\singlespace\noindent\raggedright
 This work was supported by Department of Energy
  contracts:
  DE-FG02-91ER40676 (BU),
  DE-FG03-92ER40701 (CIT),
  DE-FG03-91ER40618 (UCSB),
  DE-FG02-91ER40672 (Colorado),
  DE-FG02-91ER40677 (Illinois),
  DE-AC03-76SF00098 (LBL),
  DE-FG02-92ER40715 (Massachusetts),
  DE-AC02-76ER03069 (MIT),
  DE-FG06-85ER40224 (Oregon),
  DE-AC03-76SF00515 (SLAC),
  DE-FG05-91ER40627 (Tennessee),
  DE-AC02-76ER00881 (Wisconsin),
  DE-FG02-92ER40704 (Yale);
  National Science Foundation grants:
  PHY-91-13428 (UCSC),
  PHY-89-21320 (Columbia),
  PHY-92-04239 (Cincinnati),
  PHY-88-17930 (Rutgers),
  PHY-88-19316 (Vanderbilt),
  PHY-92-03212 (Washington);
  the UK Science and Engineering Research Council
  (Brunel and RAL);
  the Istituto Nazionale di Fisica Nucleare of Italy
  (Bologna, Ferrara, Frascati, Pisa, Padova, Perugia);
  the Natural Sciences and Engineering Research Council of Canada
  (British Columbia, Victoria, TRIUMF);
  and the Japan-US Cooperative Research Project on High Energy Physics
  (Nagoya, Tohoku).}}
 
\vsize=9in
\Frontpage
 
\vbox{\eightrm\baselineskip 11.1pt plus .35in minus .65in
  \rightline{\hbox to 1.5in{\tenrm  SLAC--PUB--6451\hfil }}
  \rightline{\hbox to 1.5in{\tenrm  March 1994\hfil }}
  \rightline{\hbox to 1.5in{\tenrm  (T/E)\hfil }}   }
 
\vskip .25truecm
 
\centerline{\fourteenpoint\bf Measurement of $\alpha_s$ from
         Energy-Energy Correlations}
\centerline{{\fourteenpoint\bf at the $Z^0$ Resonance}\doeack}
\medskip
\centerline{The SLD Collaboration$^*$}\footnote{*}{List of authors
follows the list of references.}
 
\centerline{Stanford Linear Accelerator Center}
\vskip -.05in
\centerline{Stanford University, Stanford, CA 94309}
 

\bigskip
 
\centerline{\bf ABSTRACT}
{\noindent\singlespace
We have determined the strong coupling $\as$ from a comprehensive study of
energy-energy correlations ($EEC$) and
their asymmetry ($AEEC$) in hadronic decays of $Z^0$ bosons
collected by the SLD experiment at SLAC.
The data were compared with all four available
predictions of QCD calculated up to
$\Oa2$ in perturbation theory, and also with a resummed
calculation matched to all four of these calculations.
We find large discrepancies between
$\as$ values extracted from the different $\Oa2$ calculations.
We also
find a large renormalization scale ambiguity in
$\as$ determined from the $EEC$
using the $\Oa2$ calculations;
this ambiguity is reduced in the case of the $AEEC$, and is very small
when the matched calculations are used.
Averaging over all calculations, and over the $EEC$ and $AEEC$ results,
we obtain $\asz=0.124^{+0.003}_{-0.004} (exp.) \pm 0.009 (theory).$}
 
\vfil
 
\centerline{\it Submitted to Physical Review D.}
 
\vfil
\eject
\doublespace

\parskip = 0pt
\vsize=50pc
 
\section{I. INTRODUCTION}
 
Measurement of the strong coupling $\alpha_s$ in various hard processes
and at different hard scales
is one of the crucial tests of Quantum Chromodynamics (QCD)~[\QCD].
In e$^+$e$^-$ annihilation $\as$ may be determined from inclusive
measures of the topology of hadronic events.
We have previously determined $\as$ from the
rate of multijet events in hadronic decays of $Z^0$ bosons collected
by the SLD experiment at SLAC~[\JETR].
We found $\as=0.118\pm0.002
(stat.) \pm0.003(syst.)\pm0.010 (theory)$, where the dominant
uncertainty arises from uncalculated higher order contributions in
perturbation theory.
Here we present complementary measurements of $\as$ using energy-energy
correlations ($EEC$) and the asymmetry of energy-energy correlations
($AEEC$) [\EEC].
These are
inclusive two-particle correlations that can be used to probe the
structure of hadronic events in more detail than the event
topology variables
and can be calculated perturbatively in QCD.
Comparison of $\as$ determined in this way with
that measured from event topology
variables provides a significant consistency
check of the validity of perturbative QCD.
 
The $EEC$ is defined as the normalized
energy-weighted sum over all pairs of particles whose opening
angles $\chi_{ij}$ lie between $\chi-\Delta\chi\slash 2$ and
             $\chi+\Delta\chi\slash 2$:
$$
EEC(\chi)={1\over N_{event}}
         \sum_{1}^{N_{event}}
         \Biggl({1 \over {\Delta\chi}}
         {\int_{\chi-{\Delta\chi \over 2}}
              ^{\chi+{\Delta\chi \over 2}}
         \sum_{i,j=1}^{n_{particle}}
         {{E_i E_j} \over E_{vis}^2}
         \delta(\chi^\prime-\chi_{ij}) d\chi^\prime}\Biggr), \eqno\eq
$$
where $\chi$ is an opening angle to be studied for the correlations;
$\Delta\chi$ is a bin width;
$E_i$ and $E_j$ are the energies of particles $i$ and $j$ and
$E_{vis}$ is the sum of the energies of all
particles in the event. In the central region, $\chi\sim90^\circ$, the
shape of the $EEC$ is determined by hard gluon emission; hadronization
contributions are expected to be large in the collinear
and back-to-back regions,
$\chi\sim0^\circ$ and 180$^\circ$ respectively.
The asymmetry of the $EEC$ is defined as
$$
AEEC(\chi)=EEC(\pi-\chi)-EEC(\chi). \eqno\eq
$$
 
Perturbative QCD calculations of the $EEC$
were first performed up to $\Oa2$ in 1983 by
Richards, Stirling and Ellis (RSE) [\RSE]. Since then
similar calculations have been performed by
Ali and Barreiro (AB) [\AB], Falck and Kramer (FK) [\FK],
and Kunszt and Nason (KN) [\KN].
These calculations, valid in the central region, have the general form
$$
EEC(\chi)
      = {\alpha_s(\mu^2) \over 2\pi}A(\chi)
      + \biggl( {\alpha_s(\mu^2) \over 2\pi} \biggr)^2
                 \bigl[ A(\chi)2\pi b_0\ln(\mu^2/s)
                         + B(\chi) \bigr],        \eqno\eq
$$
where, to the same order in perturbation theory,
$\alpha_s(\mu^2)$ is related to the QCD scale $\lam$ by~[\PDB]
$$
\alpha_s(\mu^2)
    = {1 \over b_0 \ln(\mu^2/\lam^2) }
      \Biggl[1-{b_1 \over b_0^2}
               {\ln\bigl[\ln(\mu^2/\lam^2)\bigr]
                             \over \ln(\mu^2/\lam^2)}
      \Biggr];        \eqno\eq
$$
$\mu$ is the renormalization scale,
often expressed in terms of the factor $f=\mu^2/s$;
$\sqrt s$ is the center-of-mass energy of the experiment;
$b_0=(33-2n_f)/12\pi$; $b_1=(153-19n_f)/29\pi^2$;
and $n_f$ is the number of active quark flavors.
Here we have assumed the definition of $\lam$ for five active flavors.
The first order coefficients $A(\chi)$ can be calculated
analytically and the second order coefficients $B(\chi)$
are calculated numerically.
The main difference between the four theoretical calculations mentioned
above is in the method
used to treat the
singularities appearing in the second order coefficients
in the angular regions $\chi\sim 0^\circ$ and $\sim 180^\circ$.
 
It is important to note that
$\Oa2$ perturbative QCD calculations do not specify
the $\mu$ value to be used for any physical observable, although
this scale ambiguity will presumably vanish if the calculation is done
to all orders in perturbation theory.
Large scale ambiguities in determinations of $\as$
using such calculations for the $EEC$ and
event variables have been reported [\JETR,\LTHREEa--\OPALa].
There is an indication that the $AEEC$ may be less sensitive to higher-order
perturbative QCD contributions than the $EEC$ [\RSE],
and may
therefore be expected to be less sensitive
to changes of renormalization scale.
One also expects {\it a priori} that non-perturbative
hadronization effects will tend to cancel in the $AEEC$.
 
Furthermore, discrepancies between the four $\Oa2$
calculations of the $EEC$ have been reported [\OPALa].
These discrepancies between calculations each
supposedly complete to $\Oa2$ are not understood, but may be interpreted
as an indication that not all $\Oa2$ terms have been included in some
or all of the calculations. In the absence of further information we
assume that all calculations are equally valid
and use them all in our determination of $\as$, taking the
spread in $\as$ values as an indication of calculation uncertainty.
 
Recently progress has been made in
perturbative QCD in the form of
`resummed' calculations of certain event shape measures in $e^+e^-$
annihilation [\NLL].
These techniques have been used [\TURNOCK,\FIORE]
to calculate the $EEC$ in the
back-to-back region,
where a large contribution from soft and collinear gluon radiation appears,
by exponentiating the leading and next-to-leading
order terms in $L\equiv\ln(1/y)$, where $y={1+\cos\chi \over 2}$,
up to all orders\footnote
{\dagger}{Earlier work on a
perturbative evaluation of the leading and next-to-leading
order terms in $L$ of the $EEC$  was performed up to two-loop level by
Collins and Soper [\COLLINS].}
in $\as$ .
Within this formalism
the cumulative $EEC$ can be written [\TURNOCK]
$$
R_{EEC}(y)
      \equiv \int_0^y EEC(y') dy'
      = \Bigl(1+\sum_{n=1}^\infty\as^n C_n\Bigr) \Sigma(y,\as)
         + \sum_{n=1}^\infty\as^n Y_n(y),   \eqno\eq
$$
where
$$
 \eqalignno {
    \ln(\Sigma(y,\as))
              &= \sum_{n=1}^\infty \as^n
                         \sum_{m=1}^{n+1}  G_{nm} \ln^m(1/y)
              &\cr
              &=(\hskip 0.15in G_{12}L^2 \hskip 0.22in
                  + \hskip 0.22in G_{11}L \hskip 0.15in )\hskip 0.15in \as
              &\cr
              &+(\hskip 0.15in G_{23}L^3 \hskip 0.24in
                  +\hskip 0.22in G_{22}L^2 \hskip 0.23in
                  +\hskip 0.2in G_{21}L\hskip 0.2in )\hskip 0.15in \as^2
              &\cr
              &+\hskip 0.15in \cdots
              &\cr
              &= \underbrace{\enskip L g_1(\as L)\enskip} \enskip
                 + \enskip\underbrace{\enskip g_2(\as L)\enskip} \enskip
                 + \enskip\underbrace{\enskip \as g_3(\as L)
                                      + \cdots\enskip};
              &(6)\cr
              &  \hskip 0.35in {\rm Leading}
                 \hskip 0.32in \hbox to 1.5in{\rm Next-to-leading}
                 \hskip 0.05in {\rm Subleading}
              &\cr }
$$
 $C_n$ and $G_{nm}$ are constants; $Y_n(y)$ are
functions which vanish as $y\to 0$; and
the functions $g_1$ and $g_2$ are the sums of the leading and next-to-leading
logarithms in $L$, respectively.
Except  for the term
proportional to $G_{21}$, which can be estimated from the $\Oa2$ calculations,
the subleading terms have not been calculated.
An analytic evaluation of the
$EEC$ singularity structure [\TURNOCK]
gives an approximate simplified form:
$$
R^{resum}_{EEC}(y) = (1+C_1 \as) \exp \bigl( L g_1(\as L) + g_2(\as L)\bigr).
                                         \eqno{(7)}
$$
This resummed calculation can be combined
with each of the $\Oa2$ calculations, also expressed in the cumulative form:
$$
R_{EEC}^{\Oa2} (y)=1+{\cal A}(y)\as+{\cal B}(y)\as^2, \eqno{(8)}
$$
where ${\cal A}$ and ${\cal B}$ are the cumulative forms of
$A$ and $B$ in Eq.~(3).
Subtracting double-counted terms yields the {\it matched} form:
$$
\eqalignno{
    R^{match}_{EEC}(y)
    &= R^{resum}_{EEC}(y) + R_{EEC}^{\Oa2} (y)
               - (C_1+G_{11}L+G_{12}L^2)\as
    &(9)  \cr
    &- \bigl[G_{22}L^2+G_{23}L^3
               + (G_{11}L+G_{12}L^2)
             \bigl(C_1+{1 \over 2}(G_{11}L+G_{12}L^2)\bigr)\bigr]\as^2.
    & \cr}
$$	
This procedure is called the `\Rmatch' scheme.
\footnote
{\dagger}{Another scheme, `{\it lnR-matching}', has also been proposed
[\NLL] to combine the
resummed and $\Oa2$ calculations for event shape variables,
but it cannot be applied reliably to the $EEC$ [\OPAL].}
Equation~(9) can be differentiated with
respect to $\chi$ and compared with the
data.
A `\mRmatch' scheme is also proposed in Ref.~[\TURNOCK], whereby
the subleading term proportional to $G_{21}$ at $\Oa2$
(Eq.~(6)) is included in the
argument of the exponential in Eq.~(9).
It can be argued that
\mRmatch \ is preferred theoretically [\WEBBER]
because the extra subleading
term at $\Oa2$ is explicitly exponentiated.
 
In this analysis we attempt to be as comprehensive as
possible and compare all four $\Oa2$ calculations
with data in the central region to determine $\as$.
We then combine the resummed calculation with all four of the fixed
order calculations,
using both the \Rmatch \ and \mRmatch \ schemes, to
make an improved determination of $\as$.
We study in particular the dependence of $\as$ on the
QCD renormalization scale, which is expected
{\it a priori} to be weaker when the
resummed calculation is used.
We compare our results with $\as$ measured from our
jet rates analysis [\JETR] and with measurements from LEP
[\LTHREEa--\OPALa,\OPAL,\DELPHIa--\LTHREE].
 
\section{II. THE SLD AND EVENT SELECTION}
 
The data used in this analysis were recorded in 1992 by
the SLC Large Detector (SLD)~[\SLD]
from electron-positron
annihilation events at the $Z^0$ resonance produced by
the SLAC Linear Collider (SLC).
A subset of the components of the SLD is used for this analysis.
Charged particles are tracked in the vertex detector (VXD)~[\VXD]
and in the Central Drift Chamber (CDC) [\SLD].
Momentum measurement is provided by a uniform axial magnetic field of 0.6~T.
Particle energies are measured in the Liquid Argon Calorimeter (LAC) [\LAC],
which contains both electromagnetic and hadronic sections, and in the
Warm Iron Calorimeter [\WIC].
In this analysis the calorimeters are used only for triggering
and event selection.
 
Three triggers were used for hadronic events,
one requiring a total LAC electromagnetic energy greater than
30 GeV, another requiring at least two well-separated tracks in
the CDC, and a
third requiring at least 8 GeV in the LAC and one track in
the CDC. A selection of
hadronic events was then made by two independent methods, one based on
the topology of energy depositions in the calorimeters,
the other on the number and topology of charged tracks
measured in the CDC.
 
The analysis presented here used charged tracks measured in the CDC
and VXD. A set of cuts was applied
to the data to select well-measured tracks and events well-contained
within the detector acceptance.
Tracks were required to have:
(1)
a distance from the measured interaction point, at the point of closest
approach, of less than 5 cm in the direction transverse to the beam axis
and 10~cm along the beam axis;
(2)
a polar angle $\theta$
with respect to the beam axis within $|{\rm cos}\theta|$ $<$ 0.80;
(3)
a momentum transverse to the beam axis of
$p_{\perp}$ $>$ 150 MeV/$c$.
Events were required to have:
(1)
a minimum of five such tracks;
(2)
a total visible energy calculated from the selected tracks of at least 20 GeV;
(3)
a thrust axis [\THRUST] direction within
$|{\rm cos}\theta_T|$ $<$ 0.71.
From our 1992 data sample 6476 events survived these cuts.
The efficiency for selecting hadronic events satisfying the
$|{\rm cos}\theta_T|$ cut is estimated to be above 96\%.
The background in the selected event sample is estimated to be
0.3$\pm$0.1\%, dominated by $Z^0 \to \tau^+ \tau^-$
events.
Distributions of single particle and event topology
observables in the selected events
are found to be well described by Monte Carlo models of
hadronic $Z^0$ boson decays~[\JETSET,\HERWIG] combined with a
simulation [\GEANT] of the SLD.
 
\section{III. MEASUREMENT OF $EEC$ AND $AEEC$}
 
The $EEC$ was calculated using all pairs of selected charged tracks,
assigning each the charged pion mass.
Figure~1 shows the measured $EEC$ and $AEEC$.
The bin width was chosen to be 3.6$^\circ$, which is much larger than the
two-particle angular resolution of the detector,
so as to minimize bin-to-bin migration effects in the data correction
procedure to be described later.
Also shown in Fig.~1 are comparisons with the JETSET [\JETSET]
and HERWIG [\HERWIG] \MC \ programs which simulate the hadronic decays
of $Z^0$ bosons, combined with a simulation of the SLD, and the same
event reconstruction and selection procedures as applied to the data.
For each program 100,000 events were generated.
For JETSET we used parameter values
determined by the TASSO Collaboration at $\sqrt{s}=35$ GeV~[\TASSO],
which have been found to be in good agreement with $Z^0$ data~[\OPALc].
For HERWIG we used the default parameter values which
were derived from comparisons with LEP data~[\WEBBERb].
Both simulations reproduce the general features of the data.
 
We used both JETSET and HERWIG simulations to
correct our data for the effects of \IR,
detector acceptance and resolution, interactions and decays
in the detector, analysis cuts,
and unmeasured neutral particles.
The measured $EEC$, $EEC_{meas}(\chi_i)$,
was corrected to the {\it hadron level}, $EEC_{hadron}(\chi_i)$,
by applying bin-by-bin correction factors, $C^{EEC}_{det}(\chi_i)$:
$$
EEC_{hadron}(\chi_i)=C^{EEC}_{det}(\chi_i)
         \times EEC_{meas}(\chi_i), \eqno{(10)}
$$
where the correction factors
were calculated by comparing \MC \ results before and after
detector simulation:
$$
C^{EEC}_{det}(\chi_i)
= { EEC_{hadron}^{MC}(\chi_i) \over EEC_{detector}^{MC}(\chi_i) },
   \eqno{(11)}
$$
where $EEC_{detector}^{MC}(\chi_i)$ represents the histogram content at
bin $\chi_i$ of the $EEC$ obtained from the charged particles of the
reconstructed \MC \ events, and
$EEC^{MC}_{hadron}(\chi_i)$ represents that from all the stable particles
with lifetimes greater than $3\times 10^{-10}$~s in the
generated \MC \ events with no initial state photon radiation.
The corrected $AEEC$ was then derived from the corrected $EEC$.
Systematic errors in this correction procedure will be discussed later.
 
We corrected the data further for the effects of hadronization
using both JETSET and HERWIG.
The {\it parton-level} $EEC$ is defined by
$$
EEC_{parton}(\chi_i)
  =C^{EEC}_{frag}(\chi_i) \times EEC_{hadron}(\chi_i), \eqno{(12)}
$$
where the correction factor $C^{EEC}_{frag}(\chi_i)$ is defined by
$$
C^{EEC}_{frag}(\chi_i)
= { EEC_{parton}^{MC}(\chi_i) \over EEC_{hadron}^{MC}(\chi_i) },
   \eqno{(13)}
$$
where $EEC_{hadron}^{MC}(\chi_i)$ is the hadron-level $EEC$ as defined
in Eq.~(11), and
$EEC_{parton}^{MC}(\chi_i)$ is the content of bin $\chi_i$
constructed at the parton level.\footnote
{\dagger}{The parton level is defined in JETSET by the parameter
$Q_0$, which was set to 1 GeV, and in HERWIG by the parameter $Q_g$,
which was set to 0.75 GeV.}
The corrected $AEEC$ was again derived from the corrected $EEC$.
 
The parton-level $EEC$ and $AEEC$ are shown in Figs.~2(a) and 3(a) respectively,
where JETSET was used for both the detector and hadronization corrections.
The correction factors for the $EEC$
for detector effects and hadronization effects
are shown in Figs.~2(b) and (c) respectively.
Figures~3(b) and (c) show the corresponding effective correction factors
for the $AEEC$,
which are defined as
$C^{AEEC}_{det}(\chi_i)=AEEC^{MC}_{hadron}(\chi_i)/AEEC^{MC}_{det}(\chi_i)$ and
$C^{AEEC}_{frag}(\chi_i)=AEEC^{MC}_{parton}(\chi_i)/AEEC^{MC}_{hadron}(\chi_i)$
respectively.
Also shown in Figs.~2(b) and (c), and Figs.~3(b) and (c) are the correction
factors calculated using HERWIG.
It is interesting to note that JETSET and HERWIG give slightly
different correction factors in the central region
of the $EEC$ (Figs.~2(b) and (c)),
although both describe the detector-level data well (Fig.~1(a)).
We have found that this is due to differences in the relative production
of charged and neutral particles, which also lead to the
small differences
seen in Fig.~1.
However, the overall corrections to the parton level are found to be
very similar in the central region.
The differences between the JETSET and HERWIG correction factors
will be discussed in the section on systematic errors (Section IV).
It is also interesting that, despite the {\it a priori} expectation
that hadronization effects should cancel in the $AEEC$, the magnitude
of the deviation of the hadronization correction factors from unity is
comparable for the $AEEC$ and $EEC$.
 
The statistical errors on the $EEC$ and $AEEC$ have strong
bin-to-bin correlations because each track in each event contributes to
several bins.
The statistical error in each bin was estimated by taking the $rms$
of the contents of that bin over ten \MC \ samples,
each with the same number of events as the data sample.
These errors were used in the fits below.
 
\section{IV. DETERMINATION OF $\alpha_s$}
 
\subsection{A. Determination of $\as$ using $\Oa2$ calculations}
 
We first determine $\as$ by comparing the four $\Oa2$ QCD calculations
with the parton-level $EEC$ and $AEEC$
derived from the data.
Each calculation was fitted to
the fully corrected measured $EEC$ and $AEEC$ by minimizing $\chi^2$
under variation of $\lam$ for fixed renormalization scale factor $f$.
The fits were restricted to the angular region
36$^\circ \le \chi \le 154.8^\circ$
for the $EEC$ and $21.6^\circ \le \chi \le 79.2^\circ$ for the $AEEC$,
the ranges being chosen in such a way that
(1)~the uncertainty on the
hadronization corrections
is smaller than 30\%,
(see Figs.~2(c) and~3(c)),
(2)~the perturbation series for the $EEC$ and the
$AEEC$ remain reasonably convergent for all four QCD calculations
[\PNBHM],
and (3)~the fit quality at $f=1$ is reasonable, $\chi^2_{dof} < 5$,
where $\chi^2_{dof}$ is the $\chi^2$ per degree of freedom of the fit.
 
Figures~2(a) and~3(a) show the results
of the four $\Oa2$ QCD fits with renormalization scale factor $f=1$,
to the $EEC$ and $AEEC$ respectively.
Figures~4(a) and 5(a) show $\asz$,
derived from the fitted $\lam$ for different values of $f$,
using the $EEC$ and $AEEC$ respectively.
The corresponding fit quality $\chi^2_{dof}$ is shown in
Figs. 4(b) and 5(b).
In order to estimate the statistical errors on $\as$
we made use of the previously generated
ten sets of \MC \ events.
We performed the same fitting procedure to the $EEC$ and the $AEEC$
for each of these sets and took the
$rms$ deviation of the ten $\as$ values thus determined
to be the statistical error of the fitted $\as$
(see Figs.~4(a) and 5(a)).
 
From Figs.~4 and 5
several features of the fit results are common to each QCD calculation:
(1) $\asz$ depends strongly on $f$, the dependence being stronger
for the $EEC$ than for the $AEEC$;
(2) at low $f$ the fit quality deteriorates rapidly, and
neither $\as$ nor its error can be interpreted meaningfully;
(3) the fits are relatively good
in the scale range $f\gsim 2 \times 10^{-3}$ for the $EEC$
and $f\gsim 5 \times 10^{-2}$ for the $AEEC$,
and there is no strong
preference for a particular scale.
Similar features were seen in our $\as$ measurement from jet rates
[\JETR].
It should be noted that
a comparison of $\as$ results from the various calculations
at any particular scale $f$ reveals systematic differences of up to
20\% in magnitude,
\footnote
{\dagger}{This has also been observed previously [\OPALa].}
 although for the $AEEC$ the three more recent
calculations
are in reasonable agreement for $f>0.01$.
 
Figures 4 and 5 represent a complete summary of our results.
However, in order
to quote a single value of $\as$ from each QCD calculation we
adopt the following procedure.
We consider the ranges $0.002\le f \le 4$ for the $EEC$ and
                       $0.01\le f \le 4$ for the $AEEC$.
The lower bounds ensure that the perturbation series for the $EEC$ and
$AEEC$ remain reasonably convergent for all four QCD calculations
[\PNBHM] and that the fit quality is reasonable.
The upper bound restricts $\mu$ to a reasonable physical region,
$\mu\le 2{\sqrt s}$.
We take the extrema of the $\as$ values in these ranges to define the scale
uncertainty.
Table 1 summarizes the measured $\as$, defined as the midpoint
of $\as$ between
the extrema, and the scale uncertainty, defined as the difference between
the extrema and the midpoint value.
 
The experimental systematic errors, which arise from uncertainties
in modeling the
acceptance, efficiency, and resolution of the detector, are
summarized in Table~2.
The largest contribution is from the understanding of the effects of our
track and event selection cuts, which we evaluate by varying the cuts
over wide ranges.
We also varied the tracking efficiency
and resolution by large amounts in our \MC \ simulations
to account for any possible biases.
In addition,
the change in $\as$ is negligible when bins at either end of the fit
range are removed.
Effects due to
limited \MC \ statistics are relatively
small compared with the other errors.
Total experimental systematic errors on $\as$ are estimated to be
${+0.002 \atop -0.003}$ for the $EEC$ and
$\pm 0.004$ for the $AEEC$.
 
The theoretical uncertainties in parton
production and hadronization were estimated
by recalculating the hadronization correction factors using JETSET
with values of $Q_0$ in the range 0.5--4.0 GeV,\footnote
{\dagger}{Differences observed for changes of the other main fragmentation
parameters in JETSET, $\sigma_q$, $a$ and $b$, are small and negligible
compared to those of $Q_0$.} and
by using HERWIG, which contains a different hadronization
model.
The uncertainties in the correction factors are shown in
Figs.~2(c) ($EEC$) and 3(c) ($AEEC$).
The estimated hadronization uncertainties in $\as$ are $\pm 0.002$
for the $EEC$ and
${+0.003 \atop -0.002}$ for the $AEEC$.\footnote
{\ddagger}{These should be compared with overall shifts in
$\alpha_s(M_Z^2)$ of $+0.014$ ($EEC$) and $-0.015$ ($AEEC$) due to the
hadronization corrections.}
 
The theoretical uncertainties from the numerical precision
of the coefficients
$A(\chi)$ and $B(\chi)$ in Eq.~(3) are found to be negligible for all
four QCD calculations.
 
Averaging over all four $\Oa2$ QCD calculations
to quote a single $\as$ value from each of the
$EEC$ and $AEEC$ analyses,
we calculated the mean and $rms$ deviation of the four $\as$ values
at each $f$ in the ranges shown in Figs.~4 and 5, respectively.   The
central value of $\as$ was taken as the midpoint
of $\as$ in that $f$~range;
the {\it calculation uncertainty} was taken as the $rms$ at
the central
value; and the {\it scale uncertainty} was taken as the difference
between the central value and the extrema.
We obtain:
$$
\eqalign{
EEC:  &  ~\alpha_s(M_Z^2)=0.127 \pm 0.001(stat.)
                     ^{+0.002} _{-0.003}(syst.)
                     \pm 0.013(theory)  \cr
AEEC: &  ~\alpha_s(M_Z^2)=0.116 \pm 0.002(stat.)
                     \pm 0.004(syst.)
                     \pm 0.006(theory).           \cr }
$$
The total theoretical uncertainty is the sum in
quadrature of contributions from
hadronization ($\pm 0.002$ for $EEC$,
               ${+0.003 \atop -0.002}$ for $AEEC$),
calculation   ($\pm$ 0.006 for $EEC$, $\pm$ 0.004 for $AEEC$), and
scale         ($\pm$ 0.011 for $EEC$, $\pm$ 0.003 for $AEEC$)
uncertainties.
These $\as$ values are in
agreement
with our measurement from jet rates [\JETR].
They are also in agreement  within experimental errors with $\as$
measurements from the $EEC$ and $AEEC$
by other groups [\LTHREEa--\OPALa,\DELPHIa].
A large theoretical uncertainty,
dominated by the scale uncertainty, was also
found in our $\as$ measurement from jet rates [\JETR] and is related
to uncalculated higher order terms in QCD perturbation theory.
The scale uncertainty for the $AEEC$ is smaller than that for the $EEC$,
perhaps indicating that the contribution from
uncalculated higher order terms is smaller in the former case
as discussed in Section I.
Finally, the discrepancy in $\as$ between the four
calculations is a significant
contribution to the total uncertainty for the $EEC$, and the
dominant contribution for the $AEEC$.
 
\subsection{B. Comparison of $\Oa2$ and $\Oa2$+resummed calculations}
 
We now compare the $\Oa2$ calculations with
the matched $\Oa2$+resummed calculations.
For illustration we use the KN $\Oa2$ calculation
and the \mRmatch \ scheme.
For this purpose we perform fits to the $EEC$ data within the angular range
90$^\circ \le \chi \le 154.8^\circ$,
where the lower limit is the kinematic limit for the resummed calculation
and the upper limit is the same as in the previous section.
Figure~6 shows the results of $\Oa2$ and matched fits,
with $f=1$,
to the parton-level corrected $EEC$.
At $f=1$, $\asz=0.126\pm0.002(stat.)$ with $\chi^2_{dof}=1.86$
for the $\Oa2$ fit, and $\asz=0.133\pm0.002(stat.)$ with $\chi^2_{dof}=0.71$
for the matched fit.
However, the $\Oa2$ calculation cannot describe the large angle region
$\chi \gsim 150^\circ$,
while the matched calculation  describes the data
even up to $\chi\sim 170^\circ$.
We show in addition in Fig.~6 a fit of the $\Oa2$ calculation with
$f=0.0033$, which results in $\asz=0.114\pm 0.001(stat)$ with
$\chi_{dof}^2=0.66$.
With this small scale choice the $\Oa2$ calculation describes the data
up to $\chi\sim 165^\circ$.
This illustrates that an $\Oa2$ calculation used with a small scale
is able to reproduce, to some extent, the effects of the resummed
logarithms in the matched calculation.
However, the small scale choice results in a smaller fitted value
of $\asz$ than the choice $f=1$.
 
Figure~7 shows the resulting $\asz$ and $\chi^2_{dof}$ values
for fits of the $\Oa2$ and matched calculations
at various values of $f$.
It is apparent that good fits to the data can only be obtained for
$f \gsim 10^{-3}$ ($\Oa2$) or $f \gsim 0.2$ (matched).\footnote
{\dagger}{Similar results were obtained in our measurement of
$\as$ from jet rates [\JETR].}
Within the range $f \gsim 0.2$ the scale dependence of the matched
calculation is significantly smaller than that of the
$\Oa2$ calculation, confirming the {\it a priori} expectation discussed
in Section~I.
 
\subsection{C. Determination of $\as$ using $\Oa2$+resummed calculations}
 
In this section
we apply both the \Rmatch \ and \mRmatch \ schemes (see Section I)
to combine the resummed
calculation with each of the four $\Oa2$ calculations of the $EEC$,
and perform fits to the data.
Note that such matched calculations cannot be applied to the $AEEC$ because
no resummed calculation is available in the forward region,
$0^\circ \le\chi\le 90^\circ$.
 
Figure 8 shows
the results of the KN $\Oa2$+resummed fits to the $EEC$,
applying \Rmatch \ and \mRmatch \
with renormalization scale factor $f=1$.
The fit range was restricted to the angular region
$90^\circ \le\chi\le 154.8^\circ$
for \Rmatch \ and 90$^\circ \le\chi\le 162^\circ$
for \mRmatch \ according
to the criteria  discussed in Section IV.A.
The \Rmatch \ case does not describe
the data in the large angle region,
$\chi \gsim 155^\circ$.
Figure~9  shows the resulting $\asz$ and $\chi^2_{dof}$ values as
a function of $f$ for \mRmatch \ with all four $\Oa2$ calculations.
The most significant features are that
$f \gsim 0.2$ is needed to fit the data, and $\as$ varies slowly with $f$,
resulting in a significantly
smaller scale uncertainty than in the $\Oa2$ case, as expected.
Once again, large differences in $\as$ are apparent between
the matched RSE, AB, FK and KN calculations.
The results for \Rmatch , shown in Fig.~10, are qualitatively similar,
but are found to have a slightly larger variation of $\as$ with $f$
than that from \mRmatch, probably
due to the neglect of the sub-leading term in the exponentiation
(Section I).
 
To quote a single value of $\as$ for each matched calculation
we adopt the same procedure as in the
$\Oa2$ fits but use the range $1/4 \le f \le 4$, where
the lower\str
\eject
 
\noindent
 bound ensures that fit qualities remain reasonable.\footnote
{\dagger}{For \Rmatch, scale factors as small as $f\sim0.06$
are allowed by the quality of the fits to the data (Fig.~10(b)).
However, Fig.~10(a) shows that the change in $\asz$
between $f=1/4$ and $f=0.06$ is at most 0.002 in magnitude
(FK calculation). The change in the mean $\asz$, averaging over all
four calculations, is smaller than 0.0004, which is negligible
compared with statistical errors.
Therefore, extending the $f$ range down to 0.06 does not
affect our final $\as$ result, and we use the range
$1/4 \le f \le 4$ for consistency with our jet rates analysis [\JETR].}
This is the same $f$~range that we considered in our $\as$ measurement from
jet rates using resummed calculations~[\JETR].
The results are shown in Table 1. We note that in each case the matched
fit yields systematically larger $\as$ values than the $\Oa2$ fit,
and that the \mRmatch \ results are systematically larger than the
\Rmatch \ results.
Taking the average over all four calculations as before,
and averaging over both matching schemes, we obtain
$$
\asz=0.130 \pm 0.002(stat.)
          {+0.002 \atop -0.003} (syst.)
          \pm 0.007 (theory),
$$
where the total theoretical uncertainty is the sum in
quadrature of contributions from
hadronization ($\pm$ 0.002),
calculation   ($\pm$ 0.005),
scale         ($\pm$ 0.002), and
matching      ($\pm$ 0.004) uncertainties.
Here we included a matching uncertainty from the difference in $\as$ values
between \Rmatch \ and \mRmatch \
as a contribution to the total theoretical uncertainty.
It is interesting to note that the dominant uncertainty in this case
arises from the discrepancies between the four $\Oa2$ calculations,
the scale uncertainty being even smaller than the total experimental error.
Resolution of this discrepancy between calculations would potentially
yield, therefore, a
precise measurement of $\as$ in e$^+$e$^-$ annihilation at the
$Z^0$ resonance. The above result is in
agreement with that from the $\Oa2$ fits to the $EEC$
within experimental errors, and with the result from the $\Oa2$
fits to the $AEEC$ within the combined experimental errors
and theoretical uncertainties.
 
\section{V. SUMMARY}
 
We have measured energy-energy correlations and their asymmetry
in hadronic $Z^0$ decays and compared our corrected data with four
$\Oa2$ perturbative QCD calculations in order to extract $\as$.
We have also combined these calculations with a resummed calculation using two
matching schemes and extracted $\as$ from the EEC.
We obtained
$$\eqalign{
\asz
&=0.127 {+0.002 \atop -0.003} (exp.) \pm 0.013(theory)%
~{\rm from}~EEC~(\Oa2 ),\cr
\asz
&=0.116 \pm 0.005(exp.) \pm 0.006(theory)%
~{\rm from}~AEEC~(\Oa2 ),\cr
\asz
&=0.130 {+0.003 \atop -0.004} (exp.) \pm 0.007(theory)%
~{\rm from}~EEC~(\Oa2 + {\rm resummed}),
}
$$
where the experimental error is the sum in quadrature of the
statistical and systematic errors.
Note that all experimental errors
and theoretical uncertainties are highly correlated.
These $\as$ values correspond to $\lam \simeq 360$MeV, 200MeV and 420MeV,
respectively.
The renormalization scale ambiguity dominates the uncertainty in the
first case, but in the second and third cases it is dominated by the
discrepancy between the four $\Oa2$ calculations.
In the third case there is also a large uncertainty from matching
the $\Oa2$ and resummed calculations.
The renormalization scale dependence of the $\Oa2$ results,
as well as the reduction of renormalization scale sensitivity
and the increase in measured $\as$ with the application of resummed
calculations, is similar to that reported in our measurement of
$\as$ from jet
rates [\JETR]. The results using $\Oa2$ and resummed+$\Oa2$
calculations are also
consistent within
experimental errors with similar analyses by the LEP experiments
[\LTHREEa--\OPALa,\OPAL,\DELPHIa--\LTHREE].
Taking an unweighted average over all three results we obtain:
$\asz=0.124^{+0.003}_{-0.004}(exp.) \pm 0.009(theory)$, corresponding to
$\lam=317^{+48}_{-64}(exp.) \pm 144(theory)$ MeV.
 
\section{Acknowledgements}
 
We thank the personnel of the SLAC accelerator department and
the technical staffs of our collaborating institutions for their
efforts which resulted in the
successful operation of the SLC and the SLD.
We thank S.J. Brodsky, S.D.~Ellis, J.W.~Gary, G.~Kramer and G.~Turnock
for helpful comments relating to this analysis.
 
\vfill\eject

\singlespace
\noindent
{\bf Table 1}:
$\asz$ and scale
uncertainties from $\Oa2$ QCD fits to the $EEC$ and the $AEEC$,
and from $\Oa2$+resummed fits to the $EEC$.
\medskip
\newbox\hdbox%
\newcount\hdrows%
\newcount\multispancount%
\newcount\ncase%
\newcount\ncols
\newcount\nrows%
\newcount\nspan%
\newcount\ntemp%
\newdimen\hdsize%
\newdimen\newhdsize%
\newdimen\parasize%
\newdimen\spreadwidth%
\newdimen\thicksize%
\newdimen\thinsize%
\newdimen\tablewidth%
\newif\ifcentertables%
\newif\ifendsize%
\newif\iffirstrow%
\newif\iftableinfo%
\newtoks\dbt%
\newtoks\hdtks%
\newtoks\savetks%
\newtoks\tableLETtokens%
\newtoks\tabletokens%
\newtoks\widthspec%
%
%
%
%
\tableinfotrue%
\catcode`\@=11
%
%
\def\tstrut{\vrule height3.1ex depth1.2ex width0pt}%
\def\and{\char`\&}
\def\tablerule{\noalign{\hrule height\thinsize depth0pt}}%
\thicksize=1.5pt
\thinsize=0.6pt
\def\thickrule{\noalign{\hrule height\thicksize depth0pt}}%
\def\ctr#1{\hfil\ #1\hfil}%
%
%
%
%
\tablewidth=-\maxdimen%
\spreadwidth=-\maxdimen%
\def\tabskipglue{0pt plus 1fil minus 1fil}%
%
%
\centertablestrue%
%
%
%
%
\parasize=4in%
\gdef\ARGS{########}
\gdef\headerARGS{####}
\def\@mpersand{&}
{\catcode`\|=13
\gdef\letbarzero{\let|0}
\gdef\letbartab{\def|{&&}}%
\gdef\letvbbar{\let\vb|}%
}
{\catcode`\&=4
\def\ampskip{&\omit\hfil&}
\catcode`\&=13
\let&0
\xdef\letampskip{\def&{\ampskip}}%
\gdef\letnovbamp{\let\novb&\let\tab&}
}
\def\begintable{
   \begingroup%
   \catcode`\|=13\letbartab\letvbbar%
   \catcode`\&=13\letampskip\letnovbamp%
   \def\multispan##1{
      \omit \mscount##1%
      \multiply\mscount\tw@\advance\mscount\m@ne%
      \loop\ifnum\mscount>\@ne \sp@n\repeat%
   }
   \def\|{%
      &\omit\widevline&%
   }%
   \ruledtable
}
\long\def\ruledtable#1\endtable{%
%
%
%
   \offinterlineskip
   \tabskip 0pt
   \def\widevline{\vrule width\thicksize}
   \def\endrow{\@mpersand\omit\hfil\crnorm\@mpersand}%
   \def\crthick{\@mpersand\crnorm\thickrule\@mpersand}%
   \def\crthickneg##1{\@mpersand\crnorm\thickrule
          \noalign{{\skip0=##1\vskip-\skip0}}\@mpersand}%
   \def\crnorule{\@mpersand\crnorm\@mpersand}%
   \def\crnoruleneg##1{\@mpersand\crnorm
          \noalign{{\skip0=##1\vskip-\skip0}}\@mpersand}%
   \let\nr=\crnorule
   \def\endtable{\@mpersand\crnorm\thickrule}%
   \let\crnorm=\cr
%
%
   \edef\cr{\@mpersand\crnorm\tablerule\@mpersand}%
   \def\crneg##1{\@mpersand\crnorm\tablerule
          \noalign{{\skip0=##1\vskip-\skip0}}\@mpersand}%
   \let\ctneg=\crthickneg
   \let\nrneg=\crnoruleneg
   \the\tableLETtokens
%
%
   \tabletokens={&#1}
%
%
   \countROWS\tabletokens\into\nrows%
   \countCOLS\tabletokens\into\ncols%
%
%
   \advance\ncols by -1%
   \divide\ncols by 2%
   \advance\nrows by 1%
%
%
   \iftableinfo %
      \immediate\write16{[Nrows=\the\nrows, Ncols=\the\ncols]}%
   \fi%
%
%
   \ifcentertables
      \ifhmode \par\fi
      \line{
      \hss
   \else %
      \hbox{%
   \fi
      \vbox{%
         \makePREAMBLE{\the\ncols}
         \edef\next{\preamble}
         \let\preamble=\next
         \makeTABLE{\preamble}{\tabletokens}
      }
      \ifcentertables \hss}\else }\fi
   \endgroup
   \tablewidth=-\maxdimen
   \spreadwidth=-\maxdimen
}
\def\makeTABLE#1#2{
   {
   \let\ifmath0
   \let\header0
   \let\multispan0
%
%
   \ncase=0%
   \ifdim\tablewidth>-\maxdimen \ncase=1\fi%
   \ifdim\spreadwidth>-\maxdimen \ncase=2\fi%
   \relax
%
   \ifcase\ncase %
      \widthspec={}%
   \or %
      \widthspec=\expandafter{\expandafter t\expandafter o%
                 \the\tablewidth}%
   \else %
      \widthspec=\expandafter{\expandafter s\expandafter p\expandafter r%
                 \expandafter e\expandafter a\expandafter d%
                 \the\spreadwidth}%
   \fi %
   \xdef\next{
      \halign\the\widthspec{%
      #1
      \noalign{\hrule height\thicksize depth0pt}
      \the#2\endtable
%
      }
   }
   }
   \next
}
\def\makePREAMBLE#1{
   \ncols=#1
   \begingroup
   \let\ARGS=0
   \edef\xtp{\widevline\ARGS\tabskip\tabskipglue%
   &\ctr{\ARGS}\tstrut}
   \advance\ncols by -1
   \loop
      \ifnum\ncols>0 %
      \advance\ncols by -1%
      \edef\xtp{\xtp&\vrule width\thinsize\ARGS&\ctr{\ARGS}}%
   \repeat
   \xdef\preamble{\xtp&\widevline\ARGS\tabskip0pt%
   \crnorm}
   \endgroup
}
\def\countROWS#1\into#2{
   \let\countREGISTER=#2%
   \countREGISTER=0%
   \expandafter\ROWcount\the#1\endcount%
}%
\def\ROWcount{%
   \afterassignment\subROWcount\let\next= %
}%
\def\subROWcount{%
   \ifx\next\endcount %
      \let\next=\relax%
   \else%
      \ncase=0%
      \ifx\next\cr %
         \global\advance\countREGISTER by 1%
         \ncase=0%
      \fi%
      \ifx\next\endrow %
         \global\advance\countREGISTER by 1%
         \ncase=0%
      \fi%
      \ifx\next\crthick %
         \global\advance\countREGISTER by 1%
         \ncase=0%
      \fi%
      \ifx\next\crnorule %
         \global\advance\countREGISTER by 1%
         \ncase=0%
      \fi%
      \ifx\next\crthickneg %
         \global\advance\countREGISTER by 1%
         \ncase=0%
      \fi%
      \ifx\next\crnoruleneg %
         \global\advance\countREGISTER by 1%
         \ncase=0%
      \fi%
      \ifx\next\crneg %
         \global\advance\countREGISTER by 1%
         \ncase=0%
      \fi%
      \ifx\next\header %
         \ncase=1%
      \fi%
      \relax%
      \ifcase\ncase %
         \let\next\ROWcount%
      \or %
         \let\next\argROWskip%
      \else %
      \fi%
   \fi%
   \next%
}
\def\counthdROWS#1\into#2{%
\dvr{10}%
   \let\countREGISTER=#2%
   \countREGISTER=0%
\dvr{11}%
\dvr{13}%
   \expandafter\hdROWcount\the#1\endcount%
\dvr{12}%
}%
\def\hdROWcount{%
   \afterassignment\subhdROWcount\let\next= %
}%
\def\subhdROWcount{%
   \ifx\next\endcount %
      \let\next=\relax%
   \else%
      \ncase=0%
      \ifx\next\cr %
         \global\advance\countREGISTER by 1%
         \ncase=0%
      \fi%
      \ifx\next\endrow %
         \global\advance\countREGISTER by 1%
         \ncase=0%
      \fi%
      \ifx\next\crthick %
         \global\advance\countREGISTER by 1%
         \ncase=0%
      \fi%
      \ifx\next\crnorule %
         \global\advance\countREGISTER by 1%
         \ncase=0%
      \fi%
      \ifx\next\header %
         \ncase=1%
      \fi%
\relax%
      \ifcase\ncase %
         \let\next\hdROWcount%
      \or%
         \let\next\arghdROWskip%
      \else %
      \fi%
   \fi%
   \next%
}%
{\catcode`\|=13\letbartab
\gdef\countCOLS#1\into#2{%
   \let\countREGISTER=#2%
   \global\countREGISTER=0%
   \global\multispancount=0%
   \global\firstrowtrue
   \expandafter\COLcount\the#1\endcount%
   \global\advance\countREGISTER by 3%
   \global\advance\countREGISTER by -\multispancount
}%
\gdef\COLcount{%
   \afterassignment\subCOLcount\let\next= %
}%
{\catcode`\&=13%
\gdef\subCOLcount{%
   \ifx\next\endcount %
      \let\next=\relax%
   \else%
      \ncase=0%
      \iffirstrow
         \ifx\next& %
            \global\advance\countREGISTER by 2%
            \ncase=0%
         \fi%
         \ifx\next\span %
            \global\advance\countREGISTER by 1%
            \ncase=0%
         \fi%
         \ifx\next| %
            \global\advance\countREGISTER by 2%
            \ncase=0%
         \fi
         \ifx\next\|
            \global\advance\countREGISTER by 2%
            \ncase=0%
         \fi
         \ifx\next\multispan
            \ncase=1%
            \global\advance\multispancount by 1%
         \fi
         \ifx\next\header
            \ncase=2%
         \fi
         \ifx\next\cr       \global\firstrowfalse \fi
         \ifx\next\endrow   \global\firstrowfalse \fi
         \ifx\next\crthick  \global\firstrowfalse \fi
         \ifx\next\crnorule \global\firstrowfalse \fi
         \ifx\next\crnoruleneg \global\firstrowfalse \fi
         \ifx\next\crthickneg  \global\firstrowfalse \fi
         \ifx\next\crneg       \global\firstrowfalse \fi
      \fi
\relax
      \ifcase\ncase %
         \let\next\COLcount%
      \or %
         \let\next\spancount%
      \or %
         \let\next\argCOLskip%
      \else %
      \fi %
   \fi%
   \next%
}%
\gdef\argROWskip#1{%
   \let\next\ROWcount \next%
}
\gdef\arghdROWskip#1{%
   \let\next\ROWcount \next%
}
\gdef\argCOLskip#1{%
   \let\next\COLcount \next%
}
}
}
\def\spancount#1{
   \nspan=#1\multiply\nspan by 2\advance\nspan by -1%
   \global\advance \countREGISTER by \nspan
   \let\next\COLcount \next}%
\def\dvr#1{\relax}%
\def\header#1{%
\dvr{1}{\let\cr=\@mpersand%
\hdtks={#1}%
\counthdROWS\hdtks\into\hdrows%
\advance\hdrows by 1%
\ifnum\hdrows=0 \hdrows=1 \fi%
\dvr{5}\makehdPREAMBLE{\the\hdrows}%
\dvr{6}\getHDdimen{#1}%
{\parindent=0pt\hsize=\hdsize{\let\ifmath0%
\xdef\next{\valign{\headerpreamble #1\crnorm}}}\dvr{7}\next\dvr{8}%
}%
}\dvr{2}}
\def\makehdPREAMBLE#1{
\dvr{3}%
\hdrows=#1
{
\let\headerARGS=0%
\let\cr=\crnorm%
\edef\xtp{\vfil\hfil\hbox{\headerARGS}\hfil\vfil}%
\advance\hdrows by -1
\loop
\ifnum\hdrows>0%
\advance\hdrows by -1%
\edef\xtp{\xtp&\vfil\hfil\hbox{\headerARGS}\hfil\vfil}%
\repeat%
\xdef\headerpreamble{\xtp\crcr}%
}
\dvr{4}}
\def\getHDdimen#1{%
\hdsize=0pt%
\getsize#1\cr\end\cr%
}
\def\getsize#1\cr{%
\endsizefalse\savetks={#1}%
\expandafter\lookend\the\savetks\cr%
\relax \ifendsize \let\next\relax \else%
\setbox\hdbox=\hbox{#1}\newhdsize=1.0\wd\hdbox%
\ifdim\newhdsize>\hdsize \hdsize=\newhdsize \fi%
\let\next\getsize \fi%
\next%
}%
\def\lookend{\afterassignment\sublookend\let\looknext= }%
\def\sublookend{\relax%
\ifx\looknext\cr %
\let\looknext\relax \else %
   \relax
   \ifx\looknext\end \global\endsizetrue \fi%
   \let\looknext=\lookend%
    \fi \looknext%
}%
%
%
\def\tablelet#1{%
   \tableLETtokens=\expandafter{\the\tableLETtokens #1}%
}%
\catcode`\@=12
\begintable
             | \multispan{4}\tstrut\hfil $\Oa2$ fits  \hfil
             | \multispan{4}\tstrut\hfil $\Oa2$+resummed fits \hfil \nr
 QCD         | \multispan{2}\tstrut\hfil $EEC$  \hfil
             | \multispan{2}\tstrut\hfil $AEEC$ \hfil
             | \multispan{4}\tstrut\hfil $EEC$  \hfil \nr
 calculation | \multispan{2}\tstrut\hfil \quad  \hfil
             | \multispan{2}\tstrut\hfil \quad  \hfil
             | \multispan{2}\tstrut\hfil mod. R-matching \hfil
             | \multispan{2}\tstrut\hfil R-matching      \hfil \nr
             | $\asz$ | ${\rm ^{Scale}_{uncertainty}}$
             | $\asz$ | ${\rm ^{Scale}_{uncertainty}}$
             | $\asz$ | ${\rm ^{Scale}_{uncertainty}}$
             | $\asz$ | ${\rm ^{Scale}_{uncertainty}}$ \cr
    AB       | 0.132  | $\pm$ 0.011      | 0.114  | $\pm$ 0.004
             | 0.138  | $\pm$ 0.001      | 0.130  | $\pm$ 0.003        \nr
    FK       | 0.119  | $\pm$ 0.013      | 0.113  | $\pm$ 0.003
             | 0.127  | $\pm$ 0.002      | 0.120  | $\pm$ 0.004        \nr
    KN       | 0.125  | $\pm$ 0.012      | 0.114  | $\pm$ 0.004
             | 0.133  | $\pm$ 0.001      | 0.124  | $\pm$ 0.004        \nr
    RSE      | 0.133  | $\pm$ 0.011      | 0.124  | $\pm$ 0.008
             | 0.140  | $\pm$ 0.001      | 0.130  | $\pm$ 0.003
\endtable
\bigskip
\noindent
{\bf Table 2}:
Summary of estimated experimental systematic errors
on $\asz$ for the $EEC$ and $AEEC$.
\medskip

\newbox\hdbox%
\newcount\hdrows%
\newcount\multispancount%
\newcount\ncase%
\newcount\ncols
\newcount\nrows%
\newcount\nspan%
\newcount\ntemp%
\newdimen\hdsize%
\newdimen\newhdsize%
\newdimen\parasize%
\newdimen\spreadwidth%
\newdimen\thicksize%
\newdimen\thinsize%
\newdimen\tablewidth%
\newif\ifcentertables%
\newif\ifendsize%
\newif\iffirstrow%
\newif\iftableinfo%
\newtoks\dbt%
\newtoks\hdtks%
\newtoks\savetks%
\newtoks\tableLETtokens%
\newtoks\tabletokens%
\newtoks\widthspec%
%
%
%
%
\tableinfotrue%
\catcode`\@=11
%
%
\def\tstrut{\vrule height3.1ex depth1.2ex width0pt}%
\def\and{\char`\&}
\def\tablerule{\noalign{\hrule height\thinsize depth0pt}}%
\thicksize=1.5pt
\thinsize=0.6pt
\def\thickrule{\noalign{\hrule height\thicksize depth0pt}}%
\def\ctr#1{\hfil\ #1\hfil}%
%
%
%
%
\tablewidth=-\maxdimen%
\spreadwidth=-\maxdimen%
\def\tabskipglue{0pt plus 1fil minus 1fil}%
%
%
\centertablestrue%
%
%
%
%
\parasize=4in%
\gdef\ARGS{########}
\gdef\headerARGS{####}
\def\@mpersand{&}
{\catcode`\|=13
\gdef\letbarzero{\let|0}
\gdef\letbartab{\def|{&&}}%
\gdef\letvbbar{\let\vb|}%
}
{\catcode`\&=4
\def\ampskip{&\omit\hfil&}
\catcode`\&=13
\let&0
\xdef\letampskip{\def&{\ampskip}}%
\gdef\letnovbamp{\let\novb&\let\tab&}
}
\def\begintable{
   \begingroup%
   \catcode`\|=13\letbartab\letvbbar%
   \catcode`\&=13\letampskip\letnovbamp%
   \def\multispan##1{
      \omit \mscount##1%
      \multiply\mscount\tw@\advance\mscount\m@ne%
      \loop\ifnum\mscount>\@ne \sp@n\repeat%
   }
   \def\|{%
      &\omit\widevline&%
   }%
   \ruledtable
}
\long\def\ruledtable#1\endtable{%
%
%
%
   \offinterlineskip
   \tabskip 0pt
   \def\widevline{\vrule width\thicksize}
   \def\endrow{\@mpersand\omit\hfil\crnorm\@mpersand}%
   \def\crthick{\@mpersand\crnorm\thickrule\@mpersand}%
   \def\crthickneg##1{\@mpersand\crnorm\thickrule
          \noalign{{\skip0=##1\vskip-\skip0}}\@mpersand}%
   \def\crnorule{\@mpersand\crnorm\@mpersand}%
   \def\crnoruleneg##1{\@mpersand\crnorm
          \noalign{{\skip0=##1\vskip-\skip0}}\@mpersand}%
   \let\nr=\crnorule
   \def\endtable{\@mpersand\crnorm\thickrule}%
   \let\crnorm=\cr
%
%
   \edef\cr{\@mpersand\crnorm\tablerule\@mpersand}%
   \def\crneg##1{\@mpersand\crnorm\tablerule
          \noalign{{\skip0=##1\vskip-\skip0}}\@mpersand}%
   \let\ctneg=\crthickneg
   \let\nrneg=\crnoruleneg
   \the\tableLETtokens
%
%
   \tabletokens={&#1}
%
%
   \countROWS\tabletokens\into\nrows%
   \countCOLS\tabletokens\into\ncols%
%
%
   \advance\ncols by -1%
   \divide\ncols by 2%
   \advance\nrows by 1%
%
%
   \iftableinfo %
      \immediate\write16{[Nrows=\the\nrows, Ncols=\the\ncols]}%
   \fi%
%
%
   \ifcentertables
      \ifhmode \par\fi
      \line{
      \hss
   \else %
      \hbox{%
   \fi
      \vbox{%
         \makePREAMBLE{\the\ncols}
         \edef\next{\preamble}
         \let\preamble=\next
         \makeTABLE{\preamble}{\tabletokens}
      }
      \ifcentertables \hss}\else }\fi
   \endgroup
   \tablewidth=-\maxdimen
   \spreadwidth=-\maxdimen
}
\def\makeTABLE#1#2{
   {
   \let\ifmath0
   \let\header0
   \let\multispan0
%
%
   \ncase=0%
   \ifdim\tablewidth>-\maxdimen \ncase=1\fi%
   \ifdim\spreadwidth>-\maxdimen \ncase=2\fi%
   \relax
%
   \ifcase\ncase %
      \widthspec={}%
   \or %
      \widthspec=\expandafter{\expandafter t\expandafter o%
                 \the\tablewidth}%
   \else %
      \widthspec=\expandafter{\expandafter s\expandafter p\expandafter r%
                 \expandafter e\expandafter a\expandafter d%
                 \the\spreadwidth}%
   \fi %
   \xdef\next{
      \halign\the\widthspec{%
      #1
      \noalign{\hrule height\thicksize depth0pt}
      \the#2\endtable
%
      }
   }
   }
   \next
}
\def\makePREAMBLE#1{
   \ncols=#1
   \begingroup
   \let\ARGS=0
   \edef\xtp{\widevline\ARGS\tabskip\tabskipglue%
   &\ctr{\ARGS}\tstrut}
   \advance\ncols by -1
   \loop
      \ifnum\ncols>0 %
      \advance\ncols by -1%
      \edef\xtp{\xtp&\vrule width\thinsize\ARGS&\ctr{\ARGS}}%
   \repeat
   \xdef\preamble{\xtp&\widevline\ARGS\tabskip0pt%
   \crnorm}
   \endgroup
}
\def\countROWS#1\into#2{
   \let\countREGISTER=#2%
   \countREGISTER=0%
   \expandafter\ROWcount\the#1\endcount%
}%
\def\ROWcount{%
   \afterassignment\subROWcount\let\next= %
}%
\def\subROWcount{%
   \ifx\next\endcount %
      \let\next=\relax%
   \else%
      \ncase=0%
      \ifx\next\cr %
         \global\advance\countREGISTER by 1%
         \ncase=0%
      \fi%
      \ifx\next\endrow %
         \global\advance\countREGISTER by 1%
         \ncase=0%
      \fi%
      \ifx\next\crthick %
         \global\advance\countREGISTER by 1%
         \ncase=0%
      \fi%
      \ifx\next\crnorule %
         \global\advance\countREGISTER by 1%
         \ncase=0%
      \fi%
      \ifx\next\crthickneg %
         \global\advance\countREGISTER by 1%
         \ncase=0%
      \fi%
      \ifx\next\crnoruleneg %
         \global\advance\countREGISTER by 1%
         \ncase=0%
      \fi%
      \ifx\next\crneg %
         \global\advance\countREGISTER by 1%
         \ncase=0%
      \fi%
      \ifx\next\header %
         \ncase=1%
      \fi%
      \relax%
      \ifcase\ncase %
         \let\next\ROWcount%
      \or %
         \let\next\argROWskip%
      \else %
      \fi%
   \fi%
   \next%
}
\def\counthdROWS#1\into#2{%
\dvr{10}%
   \let\countREGISTER=#2%
   \countREGISTER=0%
\dvr{11}%
\dvr{13}%
   \expandafter\hdROWcount\the#1\endcount%
\dvr{12}%
}%
\def\hdROWcount{%
   \afterassignment\subhdROWcount\let\next= %
}%
\def\subhdROWcount{%
   \ifx\next\endcount %
      \let\next=\relax%
   \else%
      \ncase=0%
      \ifx\next\cr %
         \global\advance\countREGISTER by 1%
         \ncase=0%
      \fi%
      \ifx\next\endrow %
         \global\advance\countREGISTER by 1%
         \ncase=0%
      \fi%
      \ifx\next\crthick %
         \global\advance\countREGISTER by 1%
         \ncase=0%
      \fi%
      \ifx\next\crnorule %
         \global\advance\countREGISTER by 1%
         \ncase=0%
      \fi%
      \ifx\next\header %
         \ncase=1%
      \fi%
\relax%
      \ifcase\ncase %
         \let\next\hdROWcount%
      \or%
         \let\next\arghdROWskip%
      \else %
      \fi%
   \fi%
   \next%
}%
{\catcode`\|=13\letbartab
\gdef\countCOLS#1\into#2{%
   \let\countREGISTER=#2%
   \global\countREGISTER=0%
   \global\multispancount=0%
   \global\firstrowtrue
   \expandafter\COLcount\the#1\endcount%
   \global\advance\countREGISTER by 3%
   \global\advance\countREGISTER by -\multispancount
}%
\gdef\COLcount{%
   \afterassignment\subCOLcount\let\next= %
}%
{\catcode`\&=13%
\gdef\subCOLcount{%
   \ifx\next\endcount %
      \let\next=\relax%
   \else%
      \ncase=0%
      \iffirstrow
         \ifx\next& %
            \global\advance\countREGISTER by 2%
            \ncase=0%
         \fi%
         \ifx\next\span %
            \global\advance\countREGISTER by 1%
            \ncase=0%
         \fi%
         \ifx\next| %
            \global\advance\countREGISTER by 2%
            \ncase=0%
         \fi
         \ifx\next\|
            \global\advance\countREGISTER by 2%
            \ncase=0%
         \fi
         \ifx\next\multispan
            \ncase=1%
            \global\advance\multispancount by 1%
         \fi
         \ifx\next\header
            \ncase=2%
         \fi
         \ifx\next\cr       \global\firstrowfalse \fi
         \ifx\next\endrow   \global\firstrowfalse \fi
         \ifx\next\crthick  \global\firstrowfalse \fi
         \ifx\next\crnorule \global\firstrowfalse \fi
         \ifx\next\crnoruleneg \global\firstrowfalse \fi
         \ifx\next\crthickneg  \global\firstrowfalse \fi
         \ifx\next\crneg       \global\firstrowfalse \fi
      \fi
\relax
      \ifcase\ncase %
         \let\next\COLcount%
      \or %
         \let\next\spancount%
      \or %
         \let\next\argCOLskip%
      \else %
      \fi %
   \fi%
   \next%
}%
\gdef\argROWskip#1{%
   \let\next\ROWcount \next%
}
\gdef\arghdROWskip#1{%
   \let\next\ROWcount \next%
}
\gdef\argCOLskip#1{%
   \let\next\COLcount \next%
}
}
}
\def\spancount#1{
   \nspan=#1\multiply\nspan by 2\advance\nspan by -1%
   \global\advance \countREGISTER by \nspan
   \let\next\COLcount \next}%
\def\dvr#1{\relax}%
\def\header#1{%
\dvr{1}{\let\cr=\@mpersand%
\hdtks={#1}%
\counthdROWS\hdtks\into\hdrows%
\advance\hdrows by 1%
\ifnum\hdrows=0 \hdrows=1 \fi%
\dvr{5}\makehdPREAMBLE{\the\hdrows}%
\dvr{6}\getHDdimen{#1}%
{\parindent=0pt\hsize=\hdsize{\let\ifmath0%
\xdef\next{\valign{\headerpreamble #1\crnorm}}}\dvr{7}\next\dvr{8}%
}%
}\dvr{2}}
\def\makehdPREAMBLE#1{
\dvr{3}%
\hdrows=#1
{
\let\headerARGS=0%
\let\cr=\crnorm%
\edef\xtp{\vfil\hfil\hbox{\headerARGS}\hfil\vfil}%
\advance\hdrows by -1
\loop
\ifnum\hdrows>0%
\advance\hdrows by -1%
\edef\xtp{\xtp&\vfil\hfil\hbox{\headerARGS}\hfil\vfil}%
\repeat%
\xdef\headerpreamble{\xtp\crcr}%
}
\dvr{4}}
\def\getHDdimen#1{%
\hdsize=0pt%
\getsize#1\cr\end\cr%
}
\def\getsize#1\cr{%
\endsizefalse\savetks={#1}%
\expandafter\lookend\the\savetks\cr%
\relax \ifendsize \let\next\relax \else%
\setbox\hdbox=\hbox{#1}\newhdsize=1.0\wd\hdbox%
\ifdim\newhdsize>\hdsize \hdsize=\newhdsize \fi%
\let\next\getsize \fi%
\next%
}%
\def\lookend{\afterassignment\sublookend\let\looknext= }%
\def\sublookend{\relax%
\ifx\looknext\cr %
\let\looknext\relax \else %
   \relax
   \ifx\looknext\end \global\endsizetrue \fi%
   \let\looknext=\lookend%
    \fi \looknext%
}%
%
%
\def\tablelet#1{%
   \tableLETtokens=\expandafter{\the\tableLETtokens #1}%
}%
\catcode`\@=12
\begintable
  Source              | $EEC$       | $AEEC$       \cr
  {\bf Tracking}      |             |              \nr
  Track selection     | ${+0.0006 \atop -0.0011}$
                      | ${+0.0006 \atop -0.0011}$  \nr
  Tracking efficiency | +0.0006
                      | +0.0013                    \nr
  Momentum resolution | -0.0010
                      | -0.0003                    \nr
  {\bf Event selection} |           |              \nr
  Multiplicity cut    | +0.0019
                      | ${+0.0026 \atop -0.0023}$  \nr
  Fiducial cut        | ${+0.0009 \atop -0.0028}$
                      | ${+0.0023 \atop -0.0024}$  \nr
  {\bf Monte Carlo statistics} | $\pm$0.0003 | $\pm$0.0007  \cr
  Total experimental error | ${+0.0023 \atop -0.0032}$
                                  | ${+0.0040 \atop -0.0042}$
\endtable
\vfill\eject
 
\refout
\bigskip\medskip
\centerline{LIST OF AUTHORS}
 
 \instfoot\iADEL
 \instfoot\iBOL
 \instfoot\iBU
 \instfoot\iBRUN
 \instfoot\iCIT
 \instfoot\iUCSB
 \instfoot\iUCSC
 \instfoot\iCIN
 \instfoot\iCOLO
 \instfoot\iCOL
 \instfoot\iFER
 \instfoot\iFRA
 \instfoot\iILL
 \instfoot\iKEK
 \instfoot\iLBL
 \instfoot\iMIT
 \instfoot\iMASS
 \instfoot\iNAG
 \instfoot\iOREG
 \instfoot\iPAD
 \instfoot\iPERU
 \instfoot\iPISA
 \instfoot\iRUT
 \instfoot\iRAL
 \instfoot\iSLAC
 \instfoot\iTENN
 \instfoot\iTOH
 \instfoot\iVAND
 \instfoot\iWASH
 \instfoot\iWISC
 \instfoot\iYALE
 %
 %
 \author{\twelvepoint\singlespace
\hbox{K. Abe                 \unskip,\iTOH}
\hbox{I. Abt                 \unskip,\iILL}
\hbox{W.W. Ash               \unskip,\iSLAC$^\dagger$}
\hbox{D. Aston               \unskip,\iSLAC}
\hbox{N. Bacchetta           \unskip,\iPAD}
\hbox{K.G. Baird             \unskip,\iRUT}
\hbox{C. Baltay              \unskip,\iYALE}
\hbox{H.R. Band              \unskip,\iWISC}
\hbox{M.B. Barakat           \unskip,\iYALE}
\hbox{G. Baranko             \unskip,\iCOLO
% ...
}
 }
 \vskip 0.3in
 %
 %
 \address{\singlespace
 \iADEL
 Adelphi University,
 Garden City, New York 11530 \break
 \iBOL
 INFN Sezione di Bologna,
 I-40126 Bologna, Italy \break
 \iBU
 Boston University,
 Boston, Massachusetts 02215 \break
 \iBRUN
 Brunel University,
 Uxbridge, Middlesex UB8 3PH, United Kingdom \break
 \iCIT
 California Institute of Technology,
 Pasadena, California 91125 \break
 \iUCSB
 University of California at Santa Barbara,
 Santa Barbara, California 93106 \break
 \iUCSC
 University of California at Santa Cruz,
 Santa Cruz, California 95064 \break
 \iCIN
 University of Cincinnati,
 Cincinnati, Ohio 45221 \break
 \iCOLO
 University of Colorado,
 Boulder, Colorado 80309 \break
 \iCOL
 Columbia University,
 New York, New York 10027 \break
 \iFER
 INFN Sezione di Ferrara and Universit\`a di Ferrara,
 I-44100 Ferrara, Italy \break
 \iFRA
 INFN  Lab. Nazionali di Frascati,
 I-00044 Frascati, Italy \break
 \iILL
 University of Illinois,
 Urbana, Illinois 61801 \break
 \iKEK
 KEK National Laboratory,
 Tsukuba-shi, Ibaraki-ken 305 Japan \break
 \iLBL
 Lawrence Berkeley Laboratory, University of California,
 Berkeley, California 94720 \break
 \iMIT
 Massachusetts Institute of Technology,
 Cambridge, Massachusetts 02139 \break
 \iMASS
 University of Massachusetts,
 Amherst, Massachusetts 01003 \break
 \iNAG
 Nagoya University,
 Chikusa-ku, Nagoya 464 Japan  \break
 \iOREG
 University of Oregon,
 Eugene, Oregon 97403 \break
 \iPAD
 INFN Sezione di Padova and Universit\`a di Padova,
 I-35100 Padova, Italy \break
 \iPERU
 INFN Sezione di Perugia and Universit\`a di Perugia,
 I-06100 Perugia, Italy \break
 \iPISA
 INFN Sezione di Pisa and Universit\`a di Pisa,
 I-56100 Pisa, Italy \break
 \iRUT
 Rutgers University,
 Piscataway, New Jersey 08855 \break
 \iRAL
 Rutherford Appleton Laboratory,
 Chilton, Didcot, Oxon OX11 0QX United Kingdom \break
 \iSLAC
 Stanford Linear Accelerator Center, Stanford University,
 Stanford, California 94309 \break
 \iTENN
 University of Tennessee,
 Knoxville, Tennessee 37996 \break
 \iTOH
 Tohoku University,
 Sendai 980 Japan \break
 \iVAND
 Vanderbilt University,
 Nashville, Tennessee 37235 \break
 \iWASH
 University of Washington,
 Seattle, Washington 98195 \break
 \iWISC
 University of Wisconsin,
 Madison, Wisconsin 53706 \break
 \iYALE
 Yale University,
 New Haven, Connecticut 06511 \break
 $^\dagger$Deceased \break
 $^{(a)}$Also at the Universit\`a di Genova \break
 $^{(b)}$Also at the Universit\`a di Perugia \break
 }
 
\vfill\eject
\centerline{\bf Figure captions}
\medskip
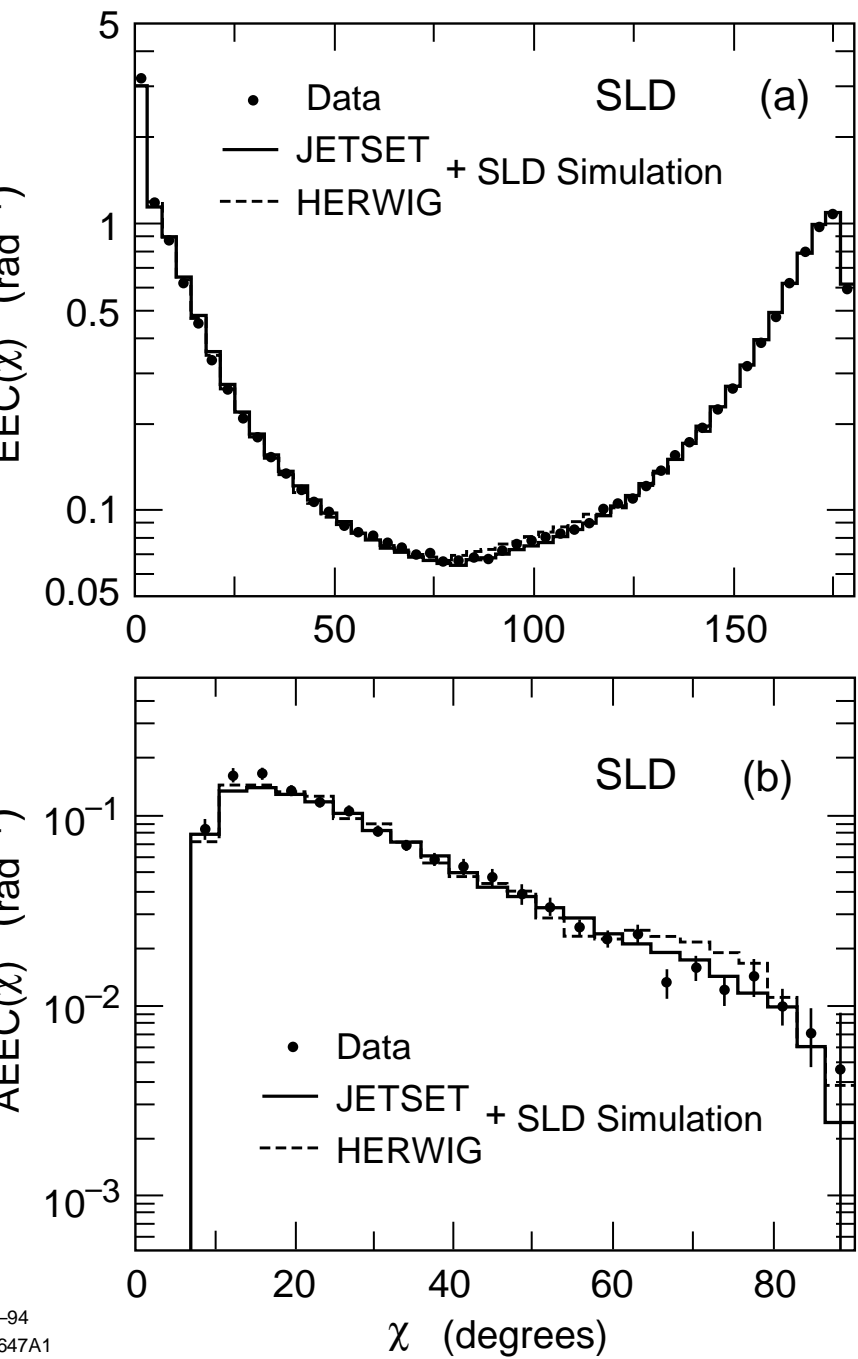
\noindent Fig.~1: \quad
          The measured (a) $EEC$ and (b) $AEEC$
          (points with error bars) compared with the JETSET (solid
          histogram) and HERWIG (dashed histogram) \MC \ simulations.
\medskip
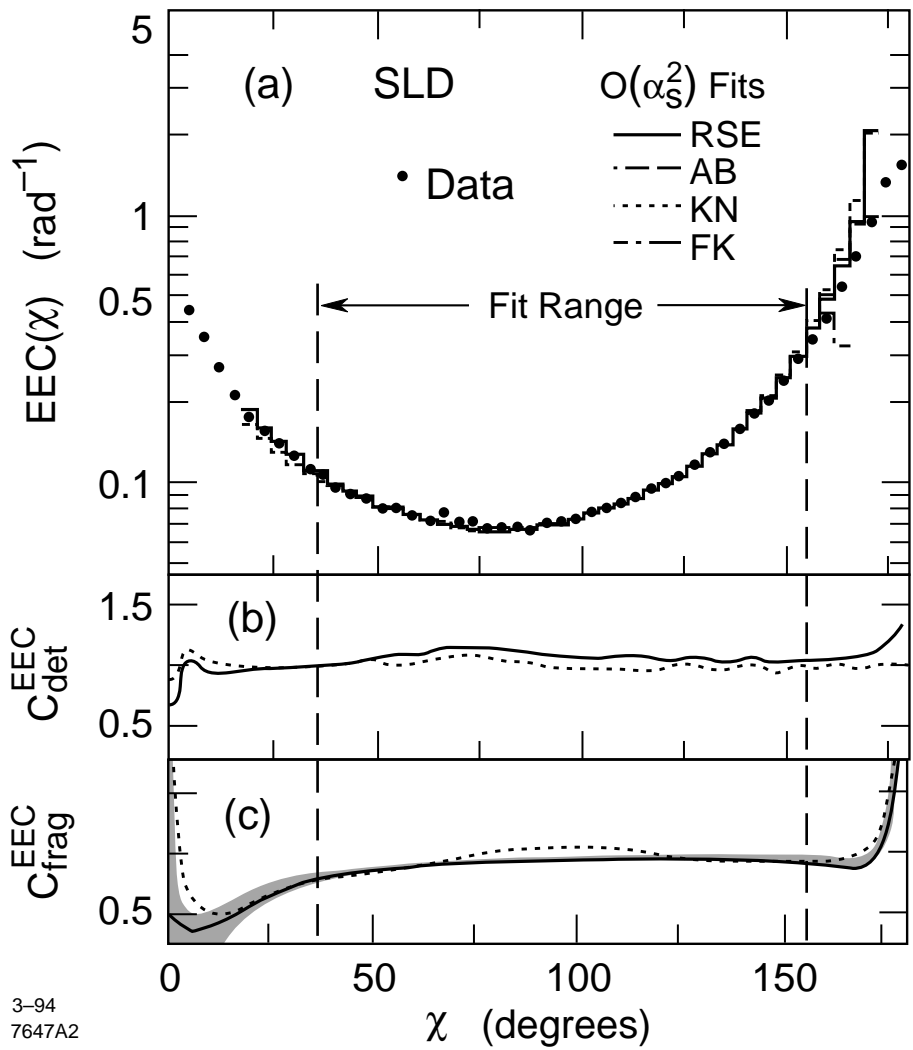
\noindent Fig.~2: \quad
          $EEC$ results: (a)  SLD data fully corrected for detector
          and hadronization effects using JETSET (see text).
          (b) Detector correction factors from JETSET (solid
          line) and HERWIG (dashed line).
          (c) Hadronization correction factors from JETSET (solid
          line with band) and HERWIG (dashed line).
          The width of the band in (c) represents the correction uncertainty
          calculated from JETSET (see text).
          Also shown as histograms in (a) are
          results of the fits of all four $\Oa2$ calculations
          at $f=1$ to the
          fully corrected $EEC$ (see text).
          The fit range is indicated by  vertical lines.
\medskip
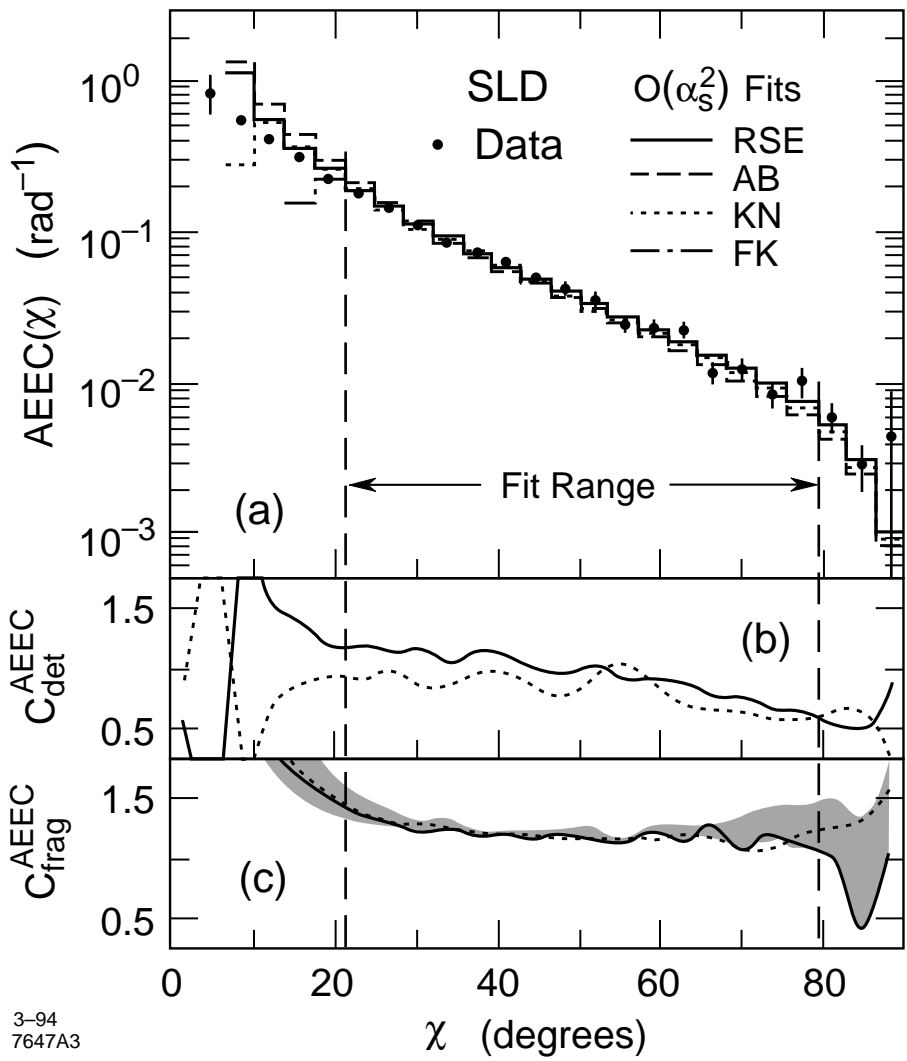
\noindent Fig.~3: \quad
          $AEEC$ results: As Fig.~2, but (b) and (c) show the
          sizes of effective correction factors (see text).
\medskip
\noindent Fig.~4: \quad
          (a) $\asz$ and (b) $\chi^2_{dof}$
          from $\Oa2$ QCD fits
          to the $EEC$ as a function of renormalization scale factor
          $f$ (see text).
          The width of the band indicates the size of statistical errors,
          shown for the KN fit only.
          The vertical lines indicate the range used in the average (see text).
\medskip
\noindent Fig.~5: \quad
          As Fig.~4, but for the $AEEC$.
\medskip
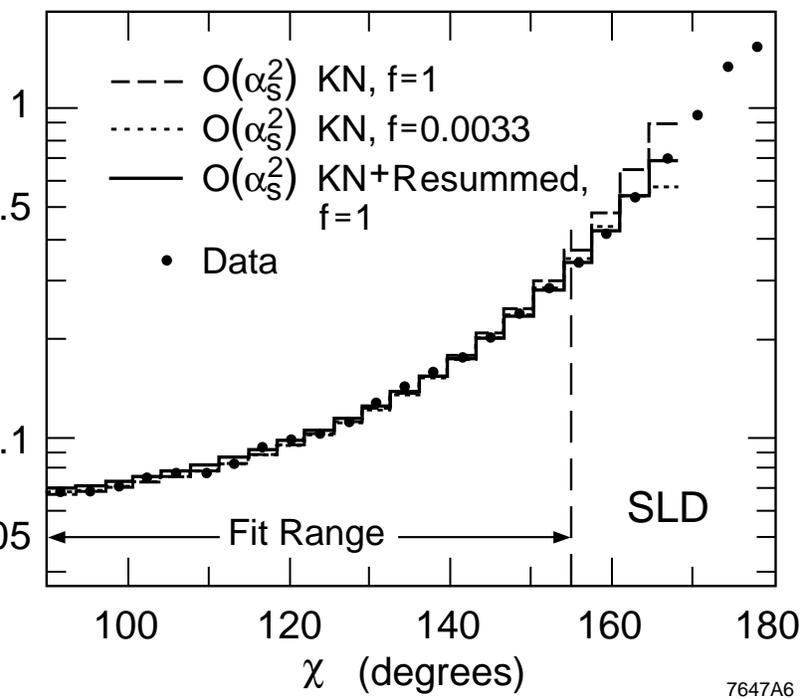
\noindent Fig.~6: \quad
          Results of the fits of KN $\Oa2$ and matched (\mRmatch)
          calculations to the
          fully corrected $EEC$. The dots represent the data points.
          The fit range is indicated by vertical lines.
          The solid histogram represents the fit of the matched calculation
          ($f=1$), the dashed histogram represents the fit
          of the $\Oa2$ calculation ($f=1$),
          and the dotted histogram represents the fit of the
          $\Oa2$ calculation ($f=0.0033$).
\medskip
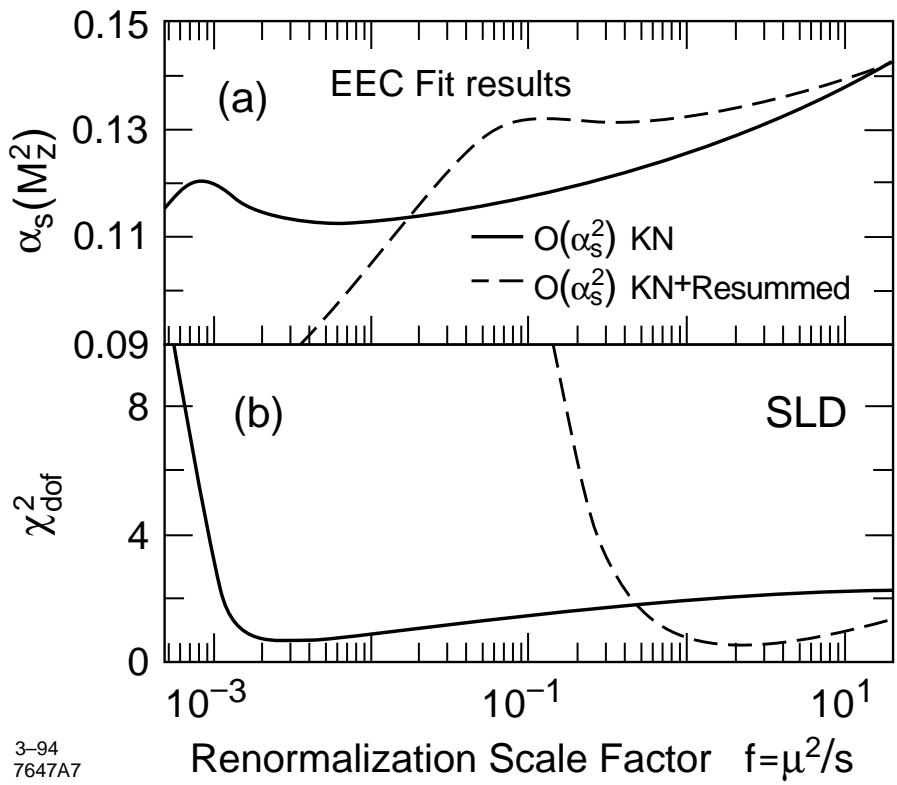
\noindent Fig.~7: \quad
          (a) $\asz$ and (b) $\chi^2_{dof}$
          from fits to the $EEC$
          using $\Oa2$ and matched (\mRmatch)
          calculations
          (see text).
\medskip
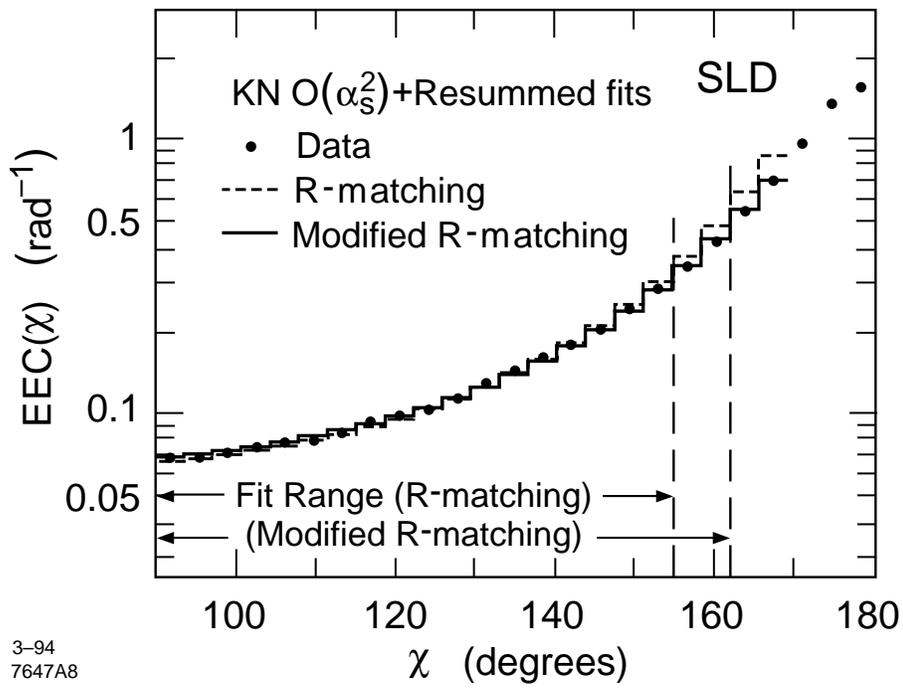
\noindent Fig.~8: \quad
          Results of the fits of the KN $\Oa2$+resummed calculation to the
          fully corrected $EEC$.
          The dots represent the data points.
	         The fit ranges are indicated by the vertical lines.
          The solid histogram (dotted histogram) represents the fit at $f=1$
          using \mRmatch \ (\Rmatch).
\medskip
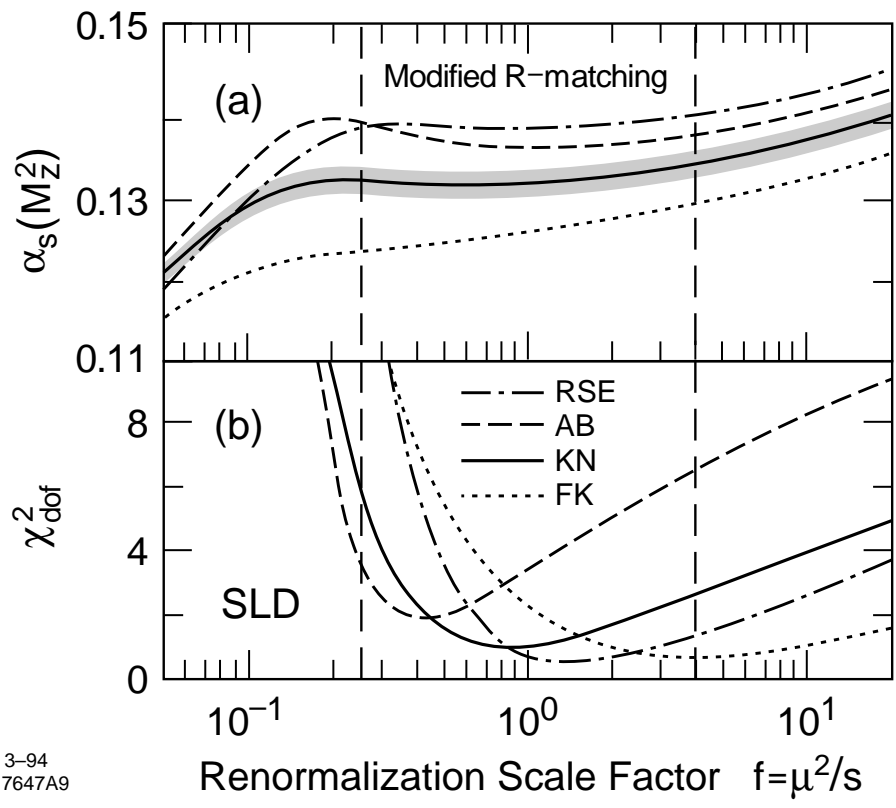
\noindent Fig.~9: \quad
          (a) $\asz$ and (b) $\chi^2_{dof}$
          from fits to the $EEC$
          using the resummed calculation combined with each of the
          four $\Oa2$ calculations using the \mRmatch \ scheme.
          The width of the band indicates the size of statistical errors,
          shown for the KN fit only.
          The vertical lines indicate the range used in the average
          (see text).
\medskip
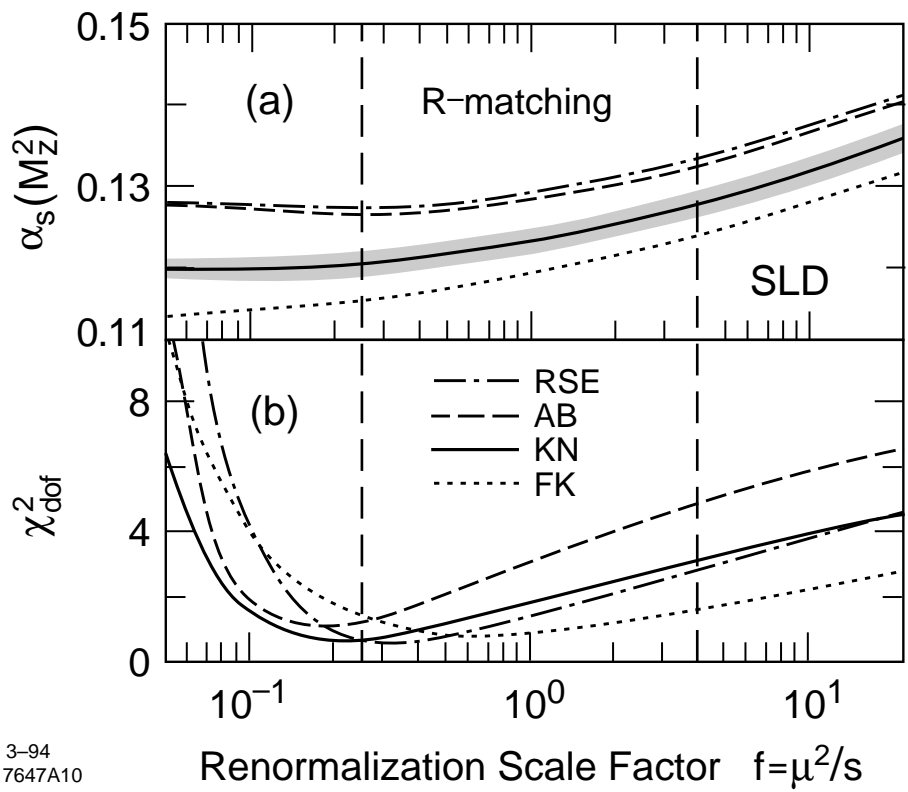
\noindent Fig.~10: \quad
          As Fig.~9, but for the \Rmatch \ scheme.
\end